\documentclass[sigconf]{acmart}
\setcopyright{none} 

\usepackage{amsmath, amsfonts, amsthm}
\usepackage[english]{babel}
\usepackage{flushend}
\usepackage{multicol}
\usepackage{color}
\usepackage{algorithm}
\usepackage{algpseudocode}
\usepackage{xspace}
\usepackage{pgfplots}
\usepackage[utf8]{inputenc}
\usepackage[T1]{fontenc}
\usepackage{sidecap}

\usepackage{float}
\newfloat{algorithm}{t}{lop}
\usepackage{booktabs}
\usepackage{subcaption}

\newtheorem{hypothesis}{Hypothesis}

\newcommand{\boldpara}[1] {\smallskip\noindent\textbf{#1---}}

\begin{document}
\title{On the (Statistical) Detection of Adversarial Examples} 

\author{Kathrin Grosse$^\dagger$, Praveen Manoharan$^\dagger$, Nicolas Papernot$^\ddagger$, Michael Backes$^{\dagger\ast}$, Patrick McDaniel$^\ddagger$\\
CISPA, Saarland Informatics Campus$^\dagger$; Penn State University$^\ddagger$;  MPI SWS$^\ast$\\
	\texttt{kathrin.grosse@cispa.saarland}}

\begin{abstract}
Machine Learning (ML) models are applied in a variety of tasks such as network
intrusion detection or malware classification. Yet, these models are vulnerable
to a class of malicious inputs known as adversarial examples. These are slightly
perturbed inputs that are classified incorrectly by the ML model. The mitigation
of these adversarial inputs remains an open problem.

As a step towards understanding adversarial examples, we show
that they are not drawn from the same distribution than the original data, and
can thus be detected using statistical tests. Using this knowledge, we introduce a complimentary approach to identify specific
inputs that are adversarial. Specifically, we augment our ML model with an additional output, in which the model is trained to classify all adversarial inputs.

We evaluate our approach\footnote{Please contact the authors to obtain the code for reproduction of the experiments.} on multiple adversarial example crafting methods (including the
fast gradient sign and saliency map methods) with several
datasets. The statistical test flags sample sets containing adversarial inputs
confidently at sample sizes between $10$ and $100$ data points. Furthermore, our augmented model either detects
adversarial examples as outliers with high accuracy ($>80$\%) or increases the adversary's
cost---the perturbation added---by more than $150$\%. In this way, we show that
statistical properties of adversarial examples are essential to their detection.\footnote{Recent work \cite{2017arXiv170507263C}, however, has shown that our approach is vulnerable if optimization-based attacks are used, which require however more computational effort.}
\end{abstract}

\maketitle

\section{Introduction}\label{section:introduction}

Machine learning algorithms are usually designed under the assumption that
models are trained on samples drawn from a distribution that is representative
of test samples for which they will later make predictions---ideally, the
training and test distributions should be identical. However, this does not hold
in the presence of adversaries. A motivated adversary may either manipulate the
training~\cite{biggio2012poisoning} or test~\cite{DBLP:conf/sp/SrndicL14}
distribution of a ML system. This has severe consequences when ML is applied to
security-critical problems. Attacks are increasingly elaborate, as demonstrated
by the variety of strategies available to evade malware detection built with
ML~\cite{DBLP:conf/sp/SrndicL14,xu2016automatically}.

Often, adversaries construct their attack inputs from a benign ML input. For
instance, the feature vector of a malware---correctly classified by a ML model
as malware---can be modified into a new feature vector, the \emph{adversarial
example}, that is classified as benign~\cite{DBLP:journals/corr/GrossePM0M16,
xu2016automatically}.

Defenses proposed to mitigate adversarial examples, such as adversarial
training~\cite{goodfellow2015explaining} and defensive
distillation~\cite{papernot2016distillation}, all fail to adapt to changes in
the attack strategy. They both make it harder for the adversary to craft
adversarial examples using existing techniques only, thus creating an arms
race~\cite{DBLP:journals/corr/CarliniW16a,papernot2016practical}. However, we
argue that this arms race is not inevitable: by definition, adversarial examples
must exhibit some statistical differences with the legitimate data on which ML
models perform well.

Hence, we develop in this work a countermeasure that uses the distinguishability
of adversarial examples with respect to the expected data distribution. We use
statistical testing to evaluate the hypothesis that adversarial examples,
crafted to evade a ML model, are outside of the training distribution. We
show that the hypothesis holds on
diverse datasets~\cite{lecun-98,arp2014drebin,RNAData} and adversarial example  algorithms~\cite{goodfellow2015explaining,papernot2016limitations,DBLP:journals/corr/PapernotMG16}.

However, this test needs to be presented with a sufficiently large
sample set of suspicious inputs---as its confidence diminishes with the number of malicious inputs in the sample set. Therefore, we propose a second complimentary mechanism for detecting
individual adversarial examples. The idea also exploits the statistical
distinguishibility of adversarial examples to design an outlier detection
system, but this time it is directly integrated in the ML model. Indeed, we show
that models can be augmented with an additional output reserved for adversarial
examples---in essence training the model with adversarial examples as their own class. The model, trained to map all adversarial inputs to the added output,
exhibits  robustness to adversaries.

Our contributions are the following:

\boldpara{Statistical Test}In Section~\ref{ssec:exp-statistical-tests}, we
employ a statistical test to distinguish adversarial examples from
the model's training data. Among tests proposed in the literature, we select the
kernel-based two-sample test introduced by Gretton et
al.~\cite{DBLP:journals/jmlr/GrettonBRSS12}. This test has the key benefit of being
model-agnostic; because its kernel allows us to apply the test directly on samples from the ML model's input data.

We demonstrate the good performance of this test on three datasets: 
MNIST (hand-written digits), DREBIN (Android malware) and MicroRNA (medical
data). Specifically, we show that the test can confidently detect samples of $50$
adversarial inputs when they differ from the expected dataset distribution. Results are consistent across multiple generation techniques for
adversarial examples, including the fast gradient sign
method~\cite{goodfellow2015explaining} and the Jacobian-based saliency map
approach~\cite{papernot2016limitations}. 

\boldpara{Integrated Outlier Detection} As the statistical test's confidence
diminishes when it is presented with increasingly small sample sets of
adversarial inputs, we propose another outlier detection system. We add an
additional class to the model's output, and train the model to recognize
adversarial examples as part of this new class. The intuition behind the idea is
the same (detect adversarial examples using their statistical properties) but
this approach allows the defender to detect individual adversarial examples
among a set of inputs identified as malicious (by the statistical test for instance). 

We observe here that adversarial examples lie in unexpected regions of the model's output surface because they are not representative of the  distribution. By training the model to identify out-of-distribution inputs, one removes at least part of its error away by filling in the things that are demonstrable (e.g., adversarial examples). 

In Section~\ref{ssec:exp-improved-models}, we find that this approach correctly
assigns adversarial examples as being part of the outlier class with over $80$\%
success for two of the three datasets considered. For the third dataset,
they are not frequently detected but the perturbation that an
adversary needs to add to mislead the model is increased by $150$\%. Thus,
the cost of conducting an attack is increased in all cases.
In addition, adversarial examples that are not detected as outliers
because they are crafted with small perturbations are often correctly classified by
our augmented model: the class of the legitimate input from which they were
generated is \emph{recovered}.

\boldpara{Arms race}We then investigate 
adversarial strategies taking into account the defense deployed. For instance, black-box attacks were previously shown to evade adversarial training and defensive distillation~\cite{papernot2016practical}. The adversary uses an  auxiliary model to find adversarial inputs that are also misclassified by the defended model (because the defended model makes adversarial crafting harder but does not solve the model error). Our
mechanisms perform well under such black-box scenarios: adversarial inputs crafted by a black-box attack are more likely to be detected than those computed directly by an
adversary with access to our model.

\section{Background}\label{section:background}
We provide here the relevant background on ML and adversarial ML. We finally
give an overview of the statistical hypothesis test applied in this paper.

\vspace*{-0.09in}

\subsection{Machine Learning Classifiers}
\label{ssec:ml-classifiers}

We introduce ML notation used throughout this paper. All ML models considered
are classifiers and learn a function $f(x) \mapsto y$. An input point or example 
$x \in X$ is made
up of $n$ components or features (e.g., all system level calls made by an
Android application), and $y \in Y$ is a label (e.g., malware or benign). In
classification problems, the possible values of $y$ are discrete. The output of
the model, however, is often real valued probabilities over the set of possible
labels, from which the most likely label is inferred as the one with the
largest probability.

In other words, there is an underlying and almost always unknown
distribution $D_\text{real}^{C_i}$ for each class $C_i$. The set of training data
$X$ is sampled from this distribution, and the classifier approximates this
distribution during training, thereby learning  $D_\text{train}^{C_i}$. The set of
test data $X_t$, used to validate the classifier's performance, is assumed to be
drawn from the same $D_\text{real}^{C_i}$.

Next, we present typical ML models used to solve classification problems and
studied in this paper.

\boldpara{Decision Trees}These models are composed of internal nodes and leafs,
whose graph makes up a tree. Each leaf is assigned one of the possible labels,
while the intermediate nodes form a path of conditions defined using the
input features. An example is classified by finding a path of appropriate
conditions from the root to one of the leaves. Decision trees are created by
successively maximizing the information gain resulting from the choice of a
condition as a way to partition the data in two subsets (according to the value
of an input feature).

\boldpara{Support Vector Machines}They compute a $n-1$ dimensional hyperplane to
separate the training points. Since there are infinitely many such hyperplanes,
the one with the largest margin is computed---yielding a convex optimization
problem given the training  data $(X,Y)$.

\boldpara{Neural Networks}They are composed of small computational
units called \emph{neurons} that apply an \emph{activation function} to their weighted 
input. Neurons are organized in interconnected layers. Depending on the number
of layers, a network is said to be \emph{shallow} (single intermediate layer) or
\emph{deep} (several intermediate layers). Information is propagated through the
network by having the output of a given layer be the input of the following
layer. Each of these links is parameterized by weights. The set of model
weights---or model parameters---are trained to minimize the model's prediction
error $\|f(x)-y\|$ on a collection of known input-output pairs $(X,Y)$.

\boldpara{Logistic regression}This linear model can be conceptualized as a
special case of neural networks without hidden layers. For problems with two
classes, the logistic function is the activation function. For multi-class
problems, it is the softmax. They are trained like neural nets.

\subsection{Adversarial Machine Learning}\label{section:adv_examples}
Adversarial ML~\cite{DBLP:conf/ccs/HuangJNRT11}, and more generally the security
and privacy of ML~\cite{papernot2016towards}, is concerned with the study of
vulnerabilities that arise when ML is deployed in the presence of malicious
individuals. Different attack vectors are available to adversaries. They can
target ML during training~\cite{biggio2012poisoning} of the model parameters or
during test time~\cite{DBLP:conf/sp/SrndicL14} when making predictions.

In this paper, we defend against test time attacks. They target a
trained model $f(\_,\theta)$, and typically aim to find an example $x'$ similar
to an original example $x$, which is however classified differently. To achieve
this, a perturbation $\delta$ with same dimensionality as $x$ is computed:
\begin{equation} \nonumber f(x',\theta)\neq f(x,\theta) \text{ where }
x'=x+\delta \text{ and }\min \delta \end{equation} where $\delta$ is chosen to
be minimal to prevent detection and as to indirectly represent the attackers
limitations when perturbing features. When targeting  computer vision, the
perturbation must not be detectable to the human eye. When targeting a malware
detector, the perturbation must not remove the application's malicious behavior.
Instead of simply having inputs classified in a wrong class, the attacker can
also target a particular class.

A typical example of such attacks is the evasion of a bayesian spam filter,
first demonstrated by Lowd et al.~\cite{DBLP:conf/ceas/LowdM05}. Malware
detection systems have also been targeted, as shown by Srndic et
al.~\cite{DBLP:conf/sp/SrndicL14} or Grosse et
al.~\cite{DBLP:journals/corr/GrossePM0M16}. In addition, these adversarial
inputs are known to transfer across (i.e., to mislead) multiple models
simultaneously~\cite{szegedy2013intriguing}. This transferability property was
used to create attacks against black-box ML systems in settings where the
adversary has no access to the model or training
data~\cite{papernot2016practical,DBLP:journals/corr/PapernotMG16}. A detailed
discussion of some of these attacks can be found in Section~\ref{sec:exp-setup}.

Several defenses for attacks at test time have been proposed. For instance,
training on adversarial inputs pro-actively~\cite{goodfellow2015explaining} or
performing defensive distillation~\cite{papernot2016distillation}. Both
of them may fail due to gradient masking~\cite{papernot2016practical}. Other
approaches make use of game theory~\cite{dalvi2004adversarial,DBLP:journals/ml/LiuC10,DBLP:conf/kdd/BrucknerS11}. However, they are
computationally expensive.

\subsection{Statistical Hypothesis Testing} \label{sec:statest} The framework of
two-sample statistical hypothesis testing was introduced to determine whether
two randomly drawn samples originate from the same distribution.

Formally, let $X\sim p$ denote that sample $X$ was drawn from a distribution
$p$. A statistical test can then be formalized as follows: let $X_1 \sim p$,
where $|X_1|=n$ and $X_2 \sim q$, where $|X_2|=m$. The null hypothesis $H_0$
states that $p=q$. The alternative hypothesis, $H_A$, on the other hand, is that
$p \neq q$. The statistical test $\mathcal{T} (X_1,X_2): \mathcal{X}^n \times
\mathcal{X}^m \to \{0,1\}$ takes both samples as its input and distinguishes
between $H_0$ and $H_A$. In particular, the p-value returned is matched to a 
significance level, denoted $\alpha$. The p-value is the probability that we 
obtain the observed outcome or a more extreme one. $\alpha$ relates to the 
confidence of the test, and an according threshold is fixed before the application of the test,  
typically at $0.05$ or $0.01$. If the p-value is smaller than the threshold, 
$H_0$ is rejected. A consistent test will reject $H_0$ when $p \neq q$
in the large sample size limit.

There are several two-sample tests for higher dimensions. For instance, the
Hotellings $T^2$ test evaluates whether two distributions have the same
mean~\cite{hotelling1931}. Several other tests depending on graph or tree
properties of the data were proposed by Friedman et
al.~\cite{Friedman_Rafsky_1979}, Rosenbaum et al.~\cite{Rosenbaum_anexact} or
Hall et al.~\cite{hall2002permutation}.

Most of these tests are not appropriate when considering data with high
dimensionality. This led Gretton et al.~\cite{DBLP:journals/jmlr/GrettonBRSS12}
to introduce a kernel-based test. 
In this case, we measure the distance between two
probabilities (represented by samples $X_1$ and $X_2$). In practice, this
distance is formalized as the biased estimator of the true Maximum Mean
Discrepancy (MMD): \begin{equation} \nonumber \text{MMD}_b [ \mathcal{F},X_1,X_2
]= \sup\limits_{f \in  \mathcal{F}}  \left(\frac{1}{n} \sum\limits_{i = 1}^n
f(x_{1i}) - \frac{1}{m} \sum\limits_{i = 1}^m f(x_{2i}) \right) \end{equation}
where the maximum indicates that we pick the kernel function $f$ from the
function class $\mathcal{F}$ that maximize the difference between the functions.
Further, in contrast to other measures, we do not need the explicit
probabilities.

Gretton et al.~\cite{DBLP:journals/jmlr/GrettonBRSS12} introduced several tests.
We focus on one of them in the following: a test based on the
asymptotic distribution of the unbiased MMD. 
However, to consistently estimate the distribution of the MMD under $H_0$, we
need to bootstrap\footnote{Other methods have been proposed, such as moment
	matching Pearson curves. We focus here on one specific test used in this paper.}.
Here, bootstrapping refers to a subsampling method, where one samples from the
data available with replacement. By repeating this procedure many times, we
obtain an estimate for the MMD value under $H_0$.


\section{Methodology}\label{section:problemsetting}
Here, we introduce a threat model to characterize the adversaries our
system is facing. We also derive a formal argument justifying the statistical
divergence of adversarial examples from benign training points. This observation
underlies the design of our defensive mechanisms.

\subsection{Threat model}

\boldpara{Adversarial knowledge} Adversarial example crafting algorithms
proposed in the literature primarily differ in the assumptions they make about
the knowledge available to adversaries~\cite{papernot2016towards}. Algorithms
fall in two classes of assumptions: \emph{white-box} and \emph{black-box}.

Adversaries operating in the white-box threat model have unfettered access to
the ML system's architecture, the value of its parameters, and its training
data. In contrast, other adversaries do not have access to this information. They
operate in a black-box threat model where they typically can interact with the
model only through an interface analog to a cryptographic oracle: it returns the
label or probability vector output by the model when presented with an input
chosen by the adversary.

In this paper, we are designing a defensive mechanism. As such, we must consider
the worst-case scenario of the strongest adversary. We therefore operate in both the
white-box and the black-box threat model. While our attacks may not be practical 
for certain ML systems, it allows us to provide stronger defensive guarantees.

\boldpara{Adversarial capabilities} These are only restricted by constraints on
the perturbations introduced to craft adversarial examples from legitimate
inputs. Such constraints vary from dataset to dataset, and as such we leave
their discussion to the description of our setup in Section~\ref{sec:exp-setup}.

\subsection{Statistical Properties of Adversarial Examples}
\newcommand{\TrueDistribution}{$D_{Ri}$ }
\newcommand{\AdversarialDistribution}{$D_{Ai}$ }
\newcommand{\LearnedDistribution}{$D_{Li}$ } When learning a classifier from
training data as described in Section~\ref{section:background}, one seeks to
learn the real distributions of features $D_\text{real}^{C_i}$ for each subset
$C_i$ corresponding to a class $i$. These subsets define a partition of the
training data, i.e. $\cup_i C_i = X$. However, due to the limited number of
training examples, any machine learning algorithm will only be able to learn an
approximation of this real distribution, the \emph{learned feature
	distributions} $D_\text{train}^{C_i}$.

A notable result in ML is that any \emph{stable} learning algorithms will learn
the real distribution $D_\text{real}^{C_i}$ up to any multiplicative factor given
a sufficient number of training examples drawn from
$D_\text{real}^{C_i}$~\cite{Bousquet02}. Stability refers here to the fact that
given a slight modification of the data, the resulting classifier and its prediction do not
change much. Coming back to our previous reasoning, however, this \emph{full
	generalization} is in practice impossible due to the finite (and often small)
number of training examples available.

The existence of adversarial examples is a manifestation of the difference
between the real feature distribution $D_\text{real}^{C_i}$ and the learned
feature distribution $D_\text{train}^{C_i}$: the adversary follows the strategy of
finding a sample drawn from $D_\text{real}^{C_i}$ that does not adhere to the
learned distribution $D_\text{train}^{C_i}$. This is only partially dependent on
the actual algorithm used to compute the adversarial example. Yet, the adversary
(or any entity as a matter of fact) does not know the real feature distribution
$D_\text{real}^{C_i}$ (otherwise one could use that distribution in lieu of the ML
model). Therefore, existing crafting algorithms generate adversarial examples by
perturbing legitimate examples drawn from $D_\text{train}^{C_i}$, as discussed 
in Section~\ref{section:adv_examples}.

Independently of how adversarial examples were generated, all adversarial
examples for a class $C_i$ will constitute a new distribution $D_\text{adv}^{C_i}$
of this class. Following the above arguments, clearly $D_\text{adv}^{C_i}$ is
consistent with $D_\text{real}^{C_i}$, since each adversarial example for a class
$C_i$ is still a data point that belongs to this class. On the other hand,
however, $D_\text{adv}^{C_i} \neq D_\text{train}^{C_i}$. This follows from a
reductio ad absurdum: if the opposite was true, adversarial examples would be
correctly classified by the classifier.

As discussed in Section~\ref{sec:statest}, consistent statistical tests can be
used to detect whether two sets or samples $X_1$ and $X_2$ were sampled from the same
distribution or not. A sufficient (possibly infinite) number of examples in each
sample allows such a consistent statistical test to detect the difference in
the distributions even if the underlying distributions of $X_1$ and $X_2$ are very similar.

Following from the above, statistical tests are natural candidates for
adversarial example detection. Adversarial examples have to inherently be
distributed differently from legitimate examples used during training. The
difference in distribution should consequently be detectable by a statistical
test. Hence, the first hypothesis we want to validate or invalidate is the ability 
of a statistical test to distinguish between benign and adversarial data points.
We have two practical limitations, one is that we can do so by observing a 
finite (and small) number of examples, the second that we are restricted to
 existing adversarial example crafting algorithms.

\begin{hypothesis}\label{hypo:stattest} We only need a bounded number of $n$
	examples to observe a measurable difference in the distribution of examples drawn
	$D_\text{adv}$ and $D_\text{train}$ using a consistent statistical test $T$.
\end{hypothesis} We validate this hypothesis in
Section~\ref{ssec:exp-statistical-tests}. We show that as few as $50$
misclassified adversarial examples per class are sufficient to observe a
measurable difference between legitimate trainings points and adversarial
examples for existing adversarial example crafting algorithms.

\subsection{Detecting Adversarial Examples}

The main limitation of statistical tests is that they cannot detect adversarial
examples on a per-input basis. Thus, the defender must be able to collect a
sufficiently large batch of adversarial inputs before it can detect the presence
of adversaries. The defender can uncover the existence of
malicious behavior (as would an intrusion detection system) but cannot identify specific inputs that
were manipulated by the adversary among batches of examples sampled (the
specific intrusion). Indeed,
sampling a single example will not allow us to confidently estimate its
distribution with a statistical test.

This may not be acceptable in security-critical applications.  A statistical
test, itself, is therefore not always suitable as a defensive mechanism.

However, we propose another approach to leverage the fact that $D_\text{adv}$ is
different from $D_\text{train}$. We augment our learning model with an additional
\emph{outlier class} $C_\text{out}$. We then train the ML model to classify
adversarial examples in that class. Technically, $C_\text{out}$ thus contains all
examples that are not drawn from any of the learned distributions
$D_\text{train}^{C_i}$. 
We seek to show that this
augmented classifier can detect newly crafted adversarial examples at test time.

\begin{hypothesis}\label{hypoo:outlierclass}
The augmented classifier with an outlier class $C_\text{out}$ successfully
detects adversarial examples. 
\end{hypothesis}

We validate Hypothesis~\ref{hypoo:outlierclass} in Section~\ref{ssec:exp-improved-models}. We also address the potential existence of an arms race between attackers and defenders in Section~\ref{section:armsRace}.


\section{Experimental setup}
\label{sec:exp-setup}

We describe here the experimental setup used in
Sections~\ref{ssec:exp-statistical-tests},~\ref{ssec:exp-improved-models}
and~\ref{section:armsRace} to validate the hypotheses stated in
Section~\ref{section:problemsetting}. Specifically, we design our setup to
answer the following experimental questions:

\begin{itemize}
\item \textbf{\emph{Q1: How well do statistical tests distinguish
adversarial distributions from legitimate ones?}} In
Section~\ref{ssec:exp-statistical-tests}, we first find that the MMD and energy
distance can statistically distinguish adversarial examples from
legitimate inputs. Statistical tests can thus be designed based on these metrics to detect adversarial examples crafted with several known techniques. In fact, we find that
often a sample size of $50$ is enough to identify them.

\item \textbf{\emph{Q2: Can detection be integrated in ML models to identify
individual adversarial examples?}} In Section~\ref{ssec:exp-improved-models},
we show that classifiers trained with an additional outlier class detects
$>80$\% of the adversarial examples it is presented with. We also find
that in cases where malicious inputs are undetected, the perturbation introduced
to evade the model needs to be increased by $150$\%, making the attack more
expensive for attackers.

\item \textbf{\emph{Q3: Do our defenses create an arms race?}}
We also find that our model with an outlier class is robust
to adaptive adversaries, such as the ones using black-box attacks. Even when such adversaries are capable of closely mimicking our model to perform
a black-box attack, they are still detected with $60$\% accuracy in the
worst case, and in many cases with accuracies larger than $>90$\%. 
\end{itemize}

To answer these questions comprehensively, we use several datasets, models
and adversarial example algorithms in an effort to represent the ML space. 
We will introduce them in more detail in the next section.

\vspace*{-0.1in}

\subsection{Adversarial example crafting}
\label{ssec:exp-adv-ex}

In our experiments, we consider the following attacks. Before we describe them, we want to remark that we do not consider functionality or utility of these attacks, in an attempt to study a worst case scenario.

\boldpara{Fast Gradient Sign Method (FGSM)}This attack computes the gradient of
the model's output with respect to its input. It then perturbs examples in that
direction. The computational efficiency of this attack comes at the expense of
it introducing large perturbations that affect the entire input. This attack is not targeted towards a particular class. We used the
initial implementation provided in the \texttt{cleverhans v.0.1}
library~\cite{DBLP:journals/corr/GoodfellowPM16} and varied the perturbation in the experiments.

\boldpara{Jacobian-based Saliency Map Approach (JSMA)}In contrast to the FGSM,
this attack iteratively computes the best feature to perturb for misclassification as a particular (usually closest) class. This yields an
adversarial example with fewer modified features, at the expense however of a
higher computational cost. We rely again on the implementation provided in the
\texttt{cleverhans v.1} library~\cite{DBLP:journals/corr/GoodfellowPM16}. 

\boldpara{SVM attack} This attack is described
in~\cite{DBLP:journals/corr/PapernotMG16}. It targets a linear SVM by shifting
the point orthogonally along the decision boundary. The result is a perturbation
similar to the one found by the FGSM. In the case of SVMs however, the
perturbation depends on the target class.

\boldpara{Decision Tree (DT) attack} We implemented a variant of the attack from~\cite{DBLP:journals/corr/PapernotMG16} where we
search the shortest path between the leaf in which the sample is currently at
and the closest leaf of another class. We then perturb the feature
that is used in the first common node shared by the two paths. By repeating
this process, we achieve a misclassification. This attack modifies only few
features and is not targeted.

\vspace*{-0.1in}

\subsection{Datasets} 
We evaluate our hypothesis on three datasets.

\boldpara{MNIST}This dataset consists of black-and-white images from $0$ to $9$
taking real values~\cite{lecun-98}. It is composed of  $60,000$ images, of which
$10,000$ form a test dataset. Each image has $28x28$ pixels.
%

\boldpara{DREBIN}This malware dataset contains $545,333$ binary malware
features~\cite{arp2014drebin}. To make adversarial example crafting faster, we
apply dimensionality reduction, as done by Grosse et
al.~\cite{DBLP:journals/corr/GrossePM0M16}, to obtain $955$ features. The
dataset contains  $129,013$ Android applications, of which $123,453$ are benign
and $5,560$ are malicious. We split this dataset randomly in training and test
data, where the test data contains one tenth of all samples.

Due to its binary nature, it is straightforward to detect attacks like the FGSM or the
SVM attack: they lead to non-binary features. We
did, nonetheless, compute them in several settings to investigate
performance of the detection capabilities. We did
not restrict the features that can be perturbed (in contrast to previous
work\cite{DBLP:journals/corr/GrossePM0M16}) in an effort to evaluate against
stronger adversaries.

\boldpara{MicroRNA}This medical dataset consists of $3966$ samples, of which $1280$ are
breast cancer serums and the remaining are non-cancer control serums. We
restrict the features to the $5$ features reported as most
useful by the original authors~\cite{RNAData}. When needed, we split the dataset randomly in training and test subsets, with a $1/10$ ratio.
This dataset contains real-valued features, each with
different mean and variance. We computed perturbations (for SVM, the FGSM and the JSMA) dependent on the variance
of the feature to be perturbed.

\vspace*{-0.1in}

\subsection{Models}
We now describe the details of the models used. These
models were already introduced in Section~\ref{ssec:ml-classifiers}, 
and their implementation available at \texttt{URL blinded}.

\boldpara{Decision Trees}We use the Gini impurity as the information gain metric to evaluate the split criterion. 

\boldpara{Support Vector Machines}We use a linear multi-class SVM. We train it with an l$2$ penalty and the squared hinge loss. When there are more then two classes, we follow the one-vs-rest strategy. 

\boldpara{Neural Networks}For MNIST, the model has two convolutional layers, with filters of size $5x5$, each followed by max pooling. A fully connected layer with $1024$ neurons follows. 

The DREBIN model reproduces the one described in~\cite{DBLP:journals/corr/GrossePM0M16}. It has two fully connected layers with $200$ neurons each. The network on the MicroRNA data has a single hidden layer with $4$ neurons.

All activation functions are ReLU. 
All models are further trained using early stopping and dropout, two common techniques to regularize the ML model's parameters and
thus improve its generalization capabilities when the model makes predictions on test data.

\boldpara{Logistic regression}We train a logistic regression on MicroRNA data with dropout and a cross-entropy loss.    

\smallskip
Most of these classifiers achieve  accuracy comparable to the state-of-the art on MNIST\footnote{The linear SVM and decision tree only achieve $92.7$\% and $67.4$\% on MNIST.}. On DREBIN, the accuracy is larger than $97.5$\%. On MicroRNA, the neural network and logistic regression achieve $95$\% accuracy. 


\section{Identifying adversarial examples using statistical metrics and tests}
\label{ssec:exp-statistical-tests}

We answer the first
question from Section~\ref{sec:exp-setup}: \emph{in practice, how well
do statistical tests distinguish adversarial distributions from legitimate
ones?} We find that two statistical metrics, the MMD and the energy distance,
both reflect changes---that adversarial examples make to the underlying
statistical properties of the distribution---by often strong variations of their
value. Armed with these metrics, we apply a statistical test. It
detects adversarial examples confidently, even when presented with
small sample sets. This validates
Hypothesis~\ref{hypo:stattest} from Section~\ref{section:problemsetting}: adversarial examples exhibit statistical properties significantly
different from legitimate data.

\subsection{Characterizing adversarial examples with statistical metrics}
\label{ssec:exp-stat-distances}
We consider two statistical distance measures commonly used to compare higher
dimensional data: (1) the maximum mean discrepancy, and (2) the energy distance.

\boldpara{Maximum Mean Discrepancy (MMD)}Recall from Section~\ref{sec:statest}
that this divergence measure is defined as:
\begin{equation}
\nonumber \text{MMD}_b [ \mathcal{F},X_1,X_2 ]= \sup\limits_{f \in  \mathcal{F}}  \left(\frac{1}{n} \sum\limits_{i = 1}^n f(x_{1i}) - \frac{1}{m} \sum\limits_{i = 1}^m f(x_{2i}) \right)
\end{equation}  
where $x_{1i} \in X_1$ is the $i$-th data point in the first sample. $x_{2j} \in X_2$ is the
$j$-th data point in the second sample, which is possibly drawn from another distribution than $X_1$. $f \in 
\mathcal{F}$ is a kernel function chosen to maximize the distances between
the samples from the two distributions. In our case, a Gaussian kernel is used.

\boldpara{Energy distance (ED)}The ED, which is also used to compare
the statistical distance between two distributions, was first introduced by
Szk\'{e}ly et al.~\cite{Szekely20131249}. It is a specific case of the maximum
mean discrepancy, where one does not apply any kernel.

\boldpara{Measurement results}We perturb the training distribution of several
MNIST models (a neural network, decision tree, and support vector machine) using
the adversarial example crafting algorithm presented in
Section~\ref{ssec:exp-adv-ex} that is suitable for each model. We then measure the statistical divergence (i.e.
the distance) between the adversarially manipulated training data and the
model's training data by computing the MMD and ED.

All data points are drawn randomly out of the $60,000$ training points. The
chance of having a particular sample and its modified counterpart in the same
batch is thus very small. To give a baseline, we also provide the distance between
the unmodified training and test distributions. At this point, we do not provide the variances, since we consider this to be a sanity check for the following steps.
We present the results of our experiments in Table~\ref{table:stapropdata}.

\begin{table}[t]
	\centering
	\begin{tabular}{@{}l rrr@{} }
	    \toprule[1.5pt]
		Manipulation & Parameters & MMD & ED \\ \midrule
		\textsl{Original} & - & \textsl{0.105} & \textsl{130.85} \\ 
		FGSM & $\varepsilon=0.07$ & 0.281  & 157.904 \\
		FGSM & $\varepsilon=0.275$ & 0.603 & 213.967\\
		JSMA & - & 0.14 & 137.63  \\
		DT attack & - & 0.1 & 130.71 \\
		SVM attack & $\epsilon=0.25$ & 0.524 & 186.32\\ 
		Flipped & - & 0.306 & 135.0 \\
		Subsampling & $45$ pixel & 2.159 & 102.7  \\
		Gaussian Blur & $4$ pixel &  1.021 & 128.52 \\
		\bottomrule[1.5pt]
	\end{tabular}
	\vspace*{2ex}
	\caption{Maximum mean discrepancy (MMD) and energy distance (ED) between the original distribution and transformed distributions obtained by several adversarial and geometric techniques on MNIST. Values are averaged over sets of $1,000$ inputs sampled randomly from the particular data. For each technique, parameters such as the perturbation magnitude for the FGSM or the number of blurred pixels are given. The JSMA leads to a change of on average $20$ pixels, whereas the DT attack changes on average $1$ pixel.}\label{table:stapropdata}
	\vspace*{-0.1in}
\end{table}

We observe that for most adversarial examples, there is a strong increase in
values of the MMD and ED. In the case of the FGSM, we observe that
the increase is stronger with larger perturbations $\epsilon$. For the JSMA and
the DT attack, changes are more subtle because these approaches only modify very
few features.

We then manipulate the test data using geometric perturbations. While these are
not adversarial, they are nevertheless helpful to interpret the magnitude of the
statistical divergences. Perturbations considered consist in mirroring
the sample, subsampling from the original values, and introducing Gaussian
blur.\footnote{This geometric perturbation approximates an attack against ML
models introduced by Biggio et al.~\cite{biggio2012poisoning}. Indeed, the
adversarial inputs produced by this attack appear as blurry, with less crisp
shapes.} We find that mirroring and subsampling affect both the MMD and
ED, whereas Gaussian blur only significantly increases the ED.

In this first experiment, we observed that there exists measurable statistical
distances between samples of benign and malicious inputs. This justifies the design of
consistent statistical tests to detect 
adversarial distributions from legitimate ones.

\vspace*{-0.1in}

\subsection{Detecting adversarial examples using hypothesis testing}
\label{ssec:exp-adv-ex-test}

We apply a statistical test to evaluate the following hypothesis: \emph{samples
from the test distribution are statistically close to samples from the training
distribution}. We expect this hypothesis to be accepted for samples from the
legitimate test distribution, but rejected for samples containing adversarial
examples. Indeed, we observed in Section~\ref{ssec:exp-stat-distances} that
adversarial distributions statistically diverge from the training distribution.

\boldpara{Two-sample hypothesis testing} 
As stated before, the test we chose is appropriate to handle high dimensional 
inputs and small sample sizes.\footnote{We used an implementation publicly available at https://github.com/emanuele/kernel\_two\_sample\_test.} We compute the biased estimate of MMD
using a Gaussian kernel, and then apply $10,000$ bootstrapping iterations 
to estimate the distributions.  
Based on this, we compute the p-value and compare it to the threshold, in our experiments $0.05$. 
For samples of legitimate data,
the observed p-value should always be very high, whereas for sample sets containing adversarial
examples, we expect it to be low---since they are sampled from a different
distribution and thus the hypothesis should be rejected.

The test is more likely to detect a
difference in two distributions when it considers samples of large size (i.e.,
the sample contains more inputs from the distribution). 

Whenever we write
\emph{confidently detected} at sample size $x$, we mean that all $200$ instances
of the test on $x$ randomly sampled examples from each of the two distributions
rejected $H_0$. Percentages reported correspond to the $n$ times of $200$ the test 
accepted $H_0$.  

\boldpara{Results} Regardless of the sample size, the hypothesis acceptance  for
benign data generally lies around $95$\%. This means that the benign data is
confidently identified as such. The sample size (i.e., the number of adversarial
examples) required to confidently detect
adversarial distributions is given in Table~\ref{table:statdatasets} for the
three datasets (MNIST, DREBIN, MicroRNA).

For most datasets and models, a sample size of $50$ adversarial examples is
sufficient for the statistical test to reject $H_0$ when comparing this sample
to a sample from the benign distribution. Thus, the statistical test identifies
adversarial examples with strong confidence, despite having few points of
comparison (relatively to the training set size of $50,000$ for MNIST).

Some exceptions should be noted. A sample of at least $100$ inputs is required
to confidently detect adversarial examples crafted with the JSMA on a neural
network. In addition, the test is unable to detect adversarial examples crafted
for MNIST on decision trees.

However, these two observations are consistent with results from
Section~\ref{ssec:exp-stat-distances}, which showed that these attacks yielded
adversarial inputs with less distinguishable statistical properties than the
FGSM for instance. Another result confirms this explanation: the SVM attack,
which was observed to lead to large changes in the MMD and energy distance, is
as well easily detected by the two-sample test. A sample of $10$ adversarial inputs
is sufficient to confidently reject the $H_0$.

\smallskip Briefly put, these results support that the distribution of JSMA or FSGM adversarial
examples differs from legitimate inputs, where we used a statistical hypothesis test
on a set of inputs.

\begin{figure}[t] 
	\begin{subfigure}[b]{\columnwidth}
			\centering
	\begin{tabular}{@{}l rrr r@{}}
	    \toprule[1.5pt]
		Dataset & FGSM & JSMA & SVM & DT \\  \midrule 
		MNIST & $50$ ($.275$) & $100$  ($16$) & $10$ ($.25$)& -  ($1$) \\
		DREBIN & $50$ ($.6$) & $50$ ($2$) & $10$ ($.25$) & $50$ ($2$) \\
		Micro & $50$ ($.6$) & $10$ ($3$) & * & $50$ ($1$) \\
		\bottomrule[1.5pt]
	\end{tabular}
			\vspace*{2ex}
			\caption{Whole datasets. The average adversarial perturbation introduced is characterized in parenthesis either by stating the perturbation parameter $\epsilon$ (FGSM, SVM attack), or the number of perturbed features (JSMA, DT attack). }
			\label{table:statdatasets}
	\end{subfigure}

	\begin{subfigure}[b]{\columnwidth}
		\centering
	\begin{tabular}{@{}l rrrrrrrr@{}}
	    \toprule[1.5pt]
		\multicolumn{1}{r}{\emph{Attack:}} & \multicolumn{2}{c}{FGSM} & \multicolumn{2}{c}{JSMA} & \multicolumn{2}{c}{SVM} & \multicolumn{2}{c}{DT} \\  
		\cmidrule(r){2-3} \cmidrule(r){4-5} \cmidrule(r){6-7}\cmidrule(l){8-9}
		\multicolumn{1}{r}{\emph{Class:}} & O & P & O & P & O & P & O & P  \\ 
        \midrule
		MNIST & $50$ & $50$ & $50$  & $100$ & $10$ & $10$ &  $50$ & - \\
		DREBIN (+) & $10$ & $10$ & $50$ & $50$ & $10$ & $10$ & $50$ & - \\
		DREBIN (-) & $10$ & $10$ & $50$ & $50$ & $10$ & $10$ & $10$ & $50$ \\
		Micro (+) & $10$ & $10$ & $10$ & $10$ & * & * & $10$ & $10$ \\ 
		Micro (-) & $10$ & $10$ & $10$ & $10$ & * & * & $10$ & $50$ \\\bottomrule[1.5pt]
	\end{tabular}
		\vspace*{2ex}
		\caption{ The statistical test is run either with the original class (O) of the input or the class predicted by the model to the perturbed input (P). Upper row for DREBIN refers to malware class (+), second to benign programs (-). For MicroRNA, (+) are the cancer serum , (-) is the control group. for MNIST, we report average values over all classes.}
		\label{table:statclasses}
	\end{subfigure}
	\caption{Minimum sample size (i.e., number of adversarial inputs) needed to confidently detect adversarial examples. Stars indicate experiments that were not conducted because the attack failed to succeed (yielding reductions in accuracy smaller than $<30$\%) or initial accuracy was too low. In other cases (-), even a sample size of $500$ was not enough to detect the adversarial examples. }
	\label{table:stat_test_combined}
\end{figure}

\begin{figure}[t]
	\centering
	\includegraphics[width=1.05\linewidth]{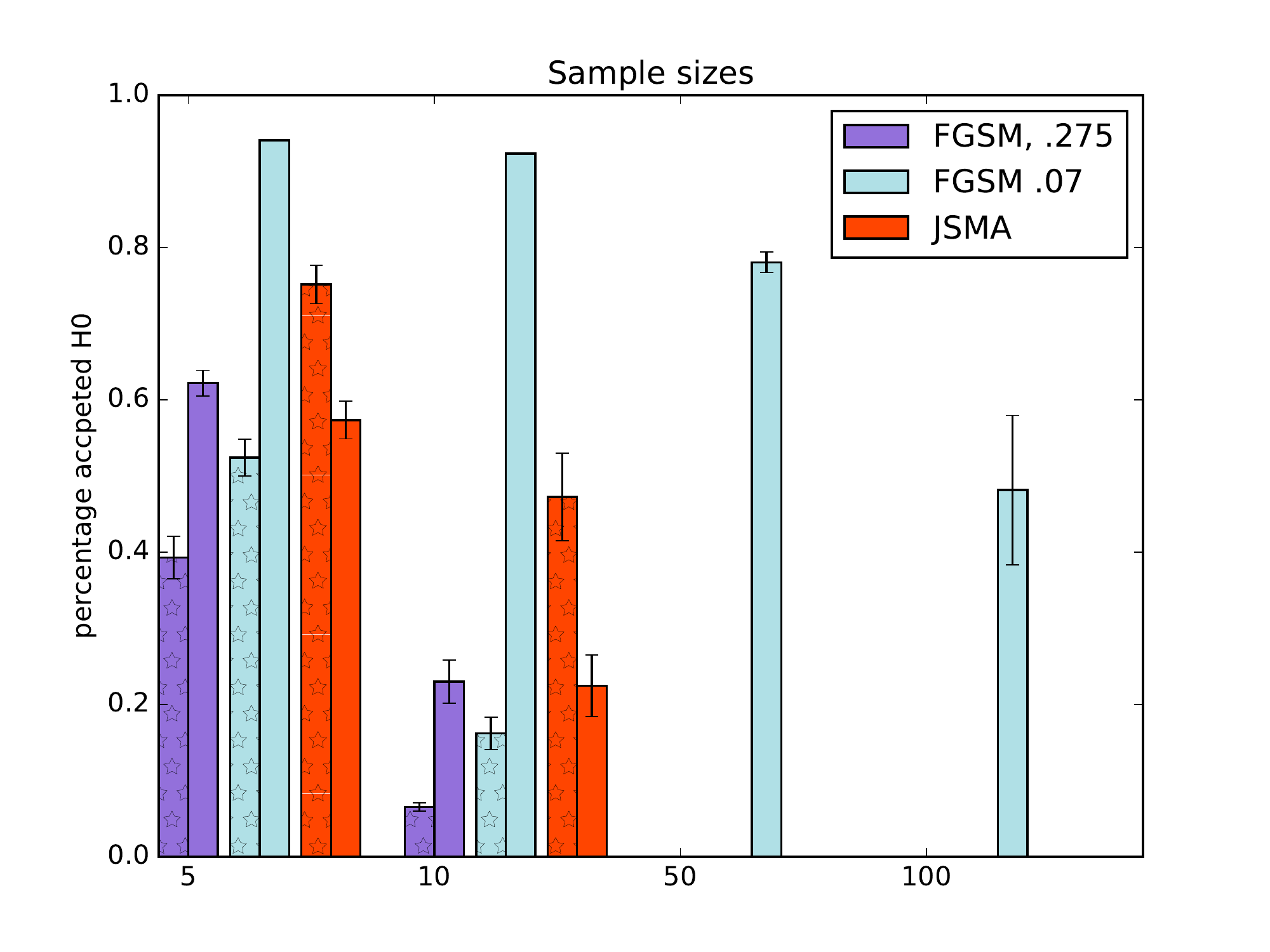} 
	\caption{Frequency of hypothesis $H_0$ acceptance with respect to the sample size (number of adversarial examples) on which the statistical test is performed. Lower values indicate that the hypothesis was rejected in more cases (e.g., the adversarial examples are detected as outside the expected distribution), which is the defender's goal. The statistical test is defined either with the original class of the input (bars with patterns), or the class predicted by the model on the perturbed input (plain bars).}
	\label{fig:sampleClass}
\end{figure}

\subsection{Decreasing sample size by performing class-wise statistical tests}
In an effort to reduce the number of inputs (i.e., the sample size) required for
the statistical test to be confident, we propose here to refine it. 
We previously assumed that the dataset was generated by a single underlying distribution. 
However, each class of the problem can be viewed as its own distribution.  
We thus perform below the statistical test on samples grouped according to their label. We separate the inputs according to their original label, or the (wrong) label assigned by the model. These tests are found to be confident for smaller sample sizes.

\boldpara{Results} Experimental results are given in Table~\ref{table:statclasses}.
On MNIST, we find that these new class-wise statistical tests reduce the sample size needed to detect the JSMA to $50$ examples. This is is also the case with DREBIN, where the sample size is reduced to $10$. For the MicroRNA we only observe a change in sample size concerning the decision tree attack, where again confident detection is already possible at a sample size of $10$.

For all datasets, we find that using a statistical test based on the distribution of the class in which the inputs are (wrongly) misclassified is more effective than using the class from which they were derived: the latter even completely fails for decision trees on MNIST. These results are consistent across the two other datasets (DREBIN and MicroRNA). In Figure~\ref{fig:sampleClass}, however, we observe that on MNIST when testing for the FGSM examples, this tendency is reversed.

Briefly put, we observed that the minimum sample size to achieve confident detection with a class-wise test is smaller than the sample size required by the general statistical test. We also noted that statistical tests comparing adversarial examples with training examples from the class they are misclassified as (rather than the class they were derived from) are more confident. 

To close this section, we want to remark that a possible conclusion from the findings in this section is to apply statistical outlier detection to detect adversarial examples. This yields a model agnostic way to detect adversarial examples. We experimented with simple outlier detection models and found, however, that many of them where not able to handle the high dimensional data with good confidence\footnote{We used the Two-Sample-Kernel Test with a single sample and Tukey's test. We further investigated several threshold/quartile based combinations for a radial, linear, and Gaussian distances and kernels.}. Since the classifiers themselves however can also be trained to perform outlier detection, we went for the approach described in the following section.

\section{Integrating Outlier Detection in Models}
\label{ssec:exp-improved-models}
In the previous section, we concluded that the distribution of adversarial
examples statistically differs from the expected distribution. 
Yet, the confidence of the test
diminishes with the number of examples in the sample set analyzed: this test
cannot be used to identify which specific inputs are adversarial among a set of
inputs.

In this Section, we provide an answer to our second experimental question: ``Can
we detect individual adversarial examples?'' Our approach adds an additional
output to the model. The  model is trained to assign this new output class to
all adversarial inputs. In other words, we explicitly train models to label all
inputs that are not part of the expected distribution as part of a new
\emph{outlier class}.

In the following  experiments, we show that this approach is complimentary to
the statistical test introduced in Section~\ref{sec:statest} because it enables
the defender to accurately identify whether a given input is adversarial or not.

\boldpara{Intuition} In the previous section, we have shown that the feature distribution of adversarial examples differs significantly 
from the distribution of benign training data. Yet, there exists no real feature distribution $D_\text{real}$ for adversarial examples: they are instead derived from the feature distribution of the original classes through minimal perturbation based on reconnaissance of the attacked classifier's behavior.

In the following, we want to leverage this insight while the classifier is being trained. Our goal is to be able to detect individual adversarial examples, as discussed in Section~\ref{section:problemsetting}. Since the distribution drift between the training and adversarial test distributions is detectable, we can hypothesize that it is learnable as well. Assuming that the classifier generalizes well to adversarial examples it has not seen during training, this would enable us to detect adversarial examples. 

\boldpara{Training with an outlier class} We start the process by training an
initial model $N_{in}$ on the original data $D=\{X,Y\}$. We compute adversarial
examples for $N_{in}$ on the training data, denoted as $X_{in}$. We then train a
new model, $N_{p1}$ on an augmented dataset, $X \cup X_{in}$, where all
adversarial examples are assigned to the outlier class. In particular, adversarial examples of different crafting algorithms are in the same class.  Specifically, we
arrange batches of inputs analyzed by the learning algorithm such that $2/3$ are
legitimate training examples from $X$ and $1/3$ are adversarial from $X_{in}$.

To evaluate the robustness of this model, we craft adversarial examples on the
test data $D_t=\{X_t,Y_t\}$ for the augmented model. In the case of JSMA, the target class will always be selected as the second most confident class, except this class is the outlier class. We thus always craft towards the strongest original class. We then report the misclassification or error on adversarial examples, e.g. all adversarial examples that were not classified as outliers or as their original class.
 Additionally, we report the percentage
of all adversarial examples that were classified as outliers.

\boldpara{Performance impact} In order for the outlier class to be a practical 
mechanism for detecting adversarial inputs, the impact on
the model accuracy on benign test data should be limited.

On MNIST, we observe that the parameters of our model augmented with an outlier
class converge slightly slower (in terms of training iterations required) to
their final values compared to an unmodified model. In addition, the model's
accuracy varies between $98.0\%$ and $98.8\%$ when it is presented with
legitimate inputs. Thus deploying the defense comes at the cost of a decrease from the original model accuracy of $99.2$\%.


On DREBIN, the original network achieves an accuracy on the benign test data of
$97.7$\%. Considering the mean, we observe a slight decrease of $0.2$\% in
accuracy for the model trained using an additional class.

On the microRNA data, the original logistic regression achieves an accuracy
around $95.4$\%. With an outlier class, accuracy drops to $93.4$\%. The same
holds for neural networks trained on the same dataset.

\boldpara{Detection results} In this series of experiments, we first measure
the detection performance of the model on malicious examples. We then study
 the failure cases of our mechanism on benign data.

Results on MNIST is given in Table~\ref{table:itattack}. 
A large proportion of adversarial examples are detected or their
original class is recovered (so they are correctly classified). 
The outlier class mechanism can be successfully trained to detect adversarial
examples produced by both the FGSM and JSMA attacks, and we discuss later in 
Section~\ref{section:armsRace} how it fairs with mixtures of both attacks.
Generally
speaking, the detection rate increases with the perturbation magnitude, while
the recovery rate decreases. 

\begin{table}
			\begin{tabular}{@{}lrrrr @{} }
				\toprule[1.5pt]
				Attack & $\epsilon$ & Recovered & Detected & Error \\ \midrule
				FGSM & $0.2$ &  $0.2$\% & $99.78$\% &  $0.02$\% \\
				FGSM & $0.28$ & $0.01$\% & $99.91$\% & $0.08$\% \\
				FGSM & $0.3$ & $0.01$\% & $99.93$\% & $0.06$\%\\
				FGSM & $0.4$ & $0$\% & $99.99$\% & $0.01$\%\\
				FGSM & $>0.5$ & $0$\%  & $100$\% & $0$\% \\
				JSMA & $\leq 80$ & $0.02$\% & $83.76$\% & $16.22$\%\\ 
                \bottomrule[1.5pt]
			\end{tabular}
				\caption{Adversarial example detection rate for models trained with an outlier class on MNIST. All models are trained on $\epsilon = 0.275$ for the FGSM and $\epsilon \leq 200$ for the JSMA. The first columns indicate the attack used upon
		completion of training and its parameter. \emph{Recovered} indicates the rate of adversarial
	examples classified in the original class of the input they were crafted from.
	\emph{Detected} indicates the percentage of adversarial examples that were
	classified as outliers. The \emph{error} rate is simply the remaining
	adversarial examples (those not correctly classified or detected as outliers).}\label{table:itattack}
\end{table}

We now report the results on the DREBIN dataset. Concerning the FGSM adversarial
examples, we observed, independently from chosen $\epsilon$, an
misclassification around $92.3$\%. The network could further not be hardened
against those adversarial examples: after training, the accuracy was still
$92.2$\%. For the JSMA, we observe an initial misclassification of $99.991$\% by
changing $2.3$ features.\footnote{This initial perturbation is much less then
reported in the original work, since we do not restrict the features as done
previously~\cite{DBLP:journals/corr/GrossePM0M16}.} When retraining on
adversarial examples, we do not observe any increase in robustness. We do
observe, however, an increase in the number of changed features up to $5.8$ when
trained on JSMA examples.

To understand whether the limited effectiveness of our defense on DREBIN is a
consequence of the binary nature of its data or the stronger success of the
attack, we implemented a second attack with a worse heuristic that initially
modifies $6.6$ features on average. By training on adversarial examples crafted
with this defense, it became impossible to craft adversarial examples using the
same modified JSMA with an upper limit of $90$ changed features.

We also trained a simple logistic regression on the MicroRNA data and trained it
on adversarial examples. The results are depicted in
Table~\ref{table:mircorna_initial}. We only applied the FGSM, since the JSMA was
not successful (only $40$\% of the adversarial examples evaded the model).
Initially, misclassification was $95.7$\% by applying a perturbation of
$\epsilon=1.0$. Further increase of $\epsilon$ had no effect. Training logistic
regression with an outlier class on adversarial examples given $\epsilon =1.0$,
we obtain a misclassification of $87.2$\%. For lower perturbations, we can
decrease misclassification, however. The limited improvement is most likely due
to the limited capacity of logistic regression models, which prevents them from
learning models robust to adversarial examples~\cite{goodfellow2015explaining}.
Thus, we trained a neural network on the data. Initially it could be evaded with
the same perturbation magnitude and success. Yet, when trained with the outlier
class, misclassification was $0.7$\% on the strongest $\epsilon$.

\begin{table}[t]
	\centering
	\begin{tabular}{@{}lrrrrr@{}}
	\toprule[1.5pt]
	    & log reg & & log reg+1 & NN+1\\
	     \cmidrule(r){2-2} \cmidrule(r){3-4} \cmidrule(r){5-5}
		 $\epsilon$ & Accuracy & Error &  Detected & Error  \\ \midrule 
		 $0.2$ & $87.4$\% &  $3.5$\% & $16.7$\% & $4.5$\% \\
		 $0.4$ & $42.9$\% &  $5.9$\% & $64.6$\% & $10.4$\% \\
		 $0.6$ & $15.9$\% &  $32.3$\% & $64.9$\% & $4.5$\% \\
	     $0.8$ & $5.1$\% & $69.2$\% & $26.3$\% & $2$\% \\ 
	    	 $1.0$ & $4.3$\% &  $87.2$\% & $6.7$\% & $0.7$\% \\
      	\bottomrule[1.5pt]
	\end{tabular}
	\vspace*{2ex}
	\caption{Accuracy and detection rate of MicroRNA logistic regressions (log reg) and neural networks (NN). We present a baseline model and two models trained with additional class (+1), both trained on FGSM at $\epsilon=1.0$ and $\epsilon=0.8$. Attack parameter for FGSM is given in the first collum. Error refers to the percentage of misclassified adversarial examples.}\label{table:mircorna_initial}
\end{table}

\begin{figure}[t]
  \centering
  \includegraphics[width=.49\linewidth]{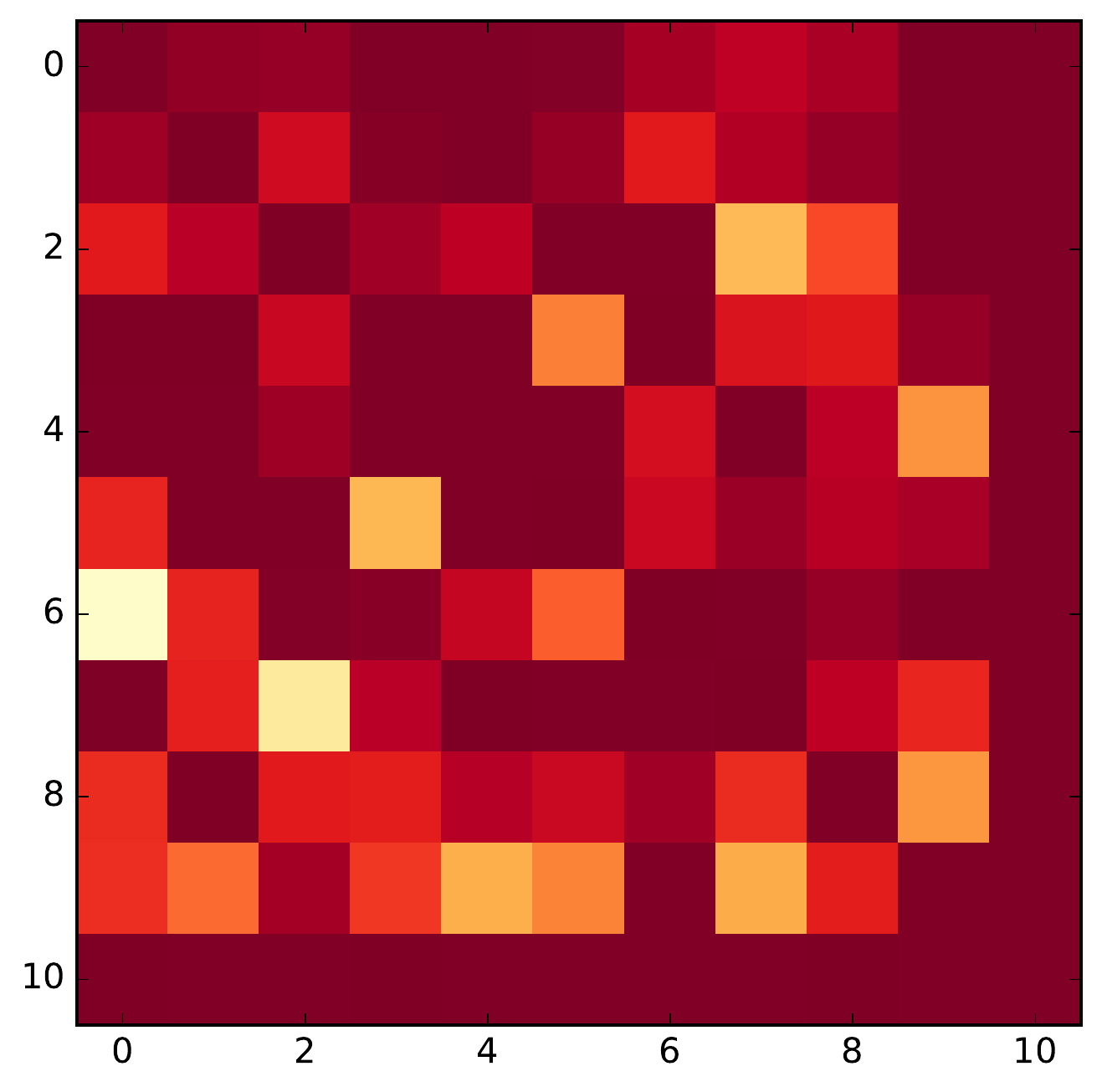}
  \centering
  \includegraphics[width=.49\linewidth]{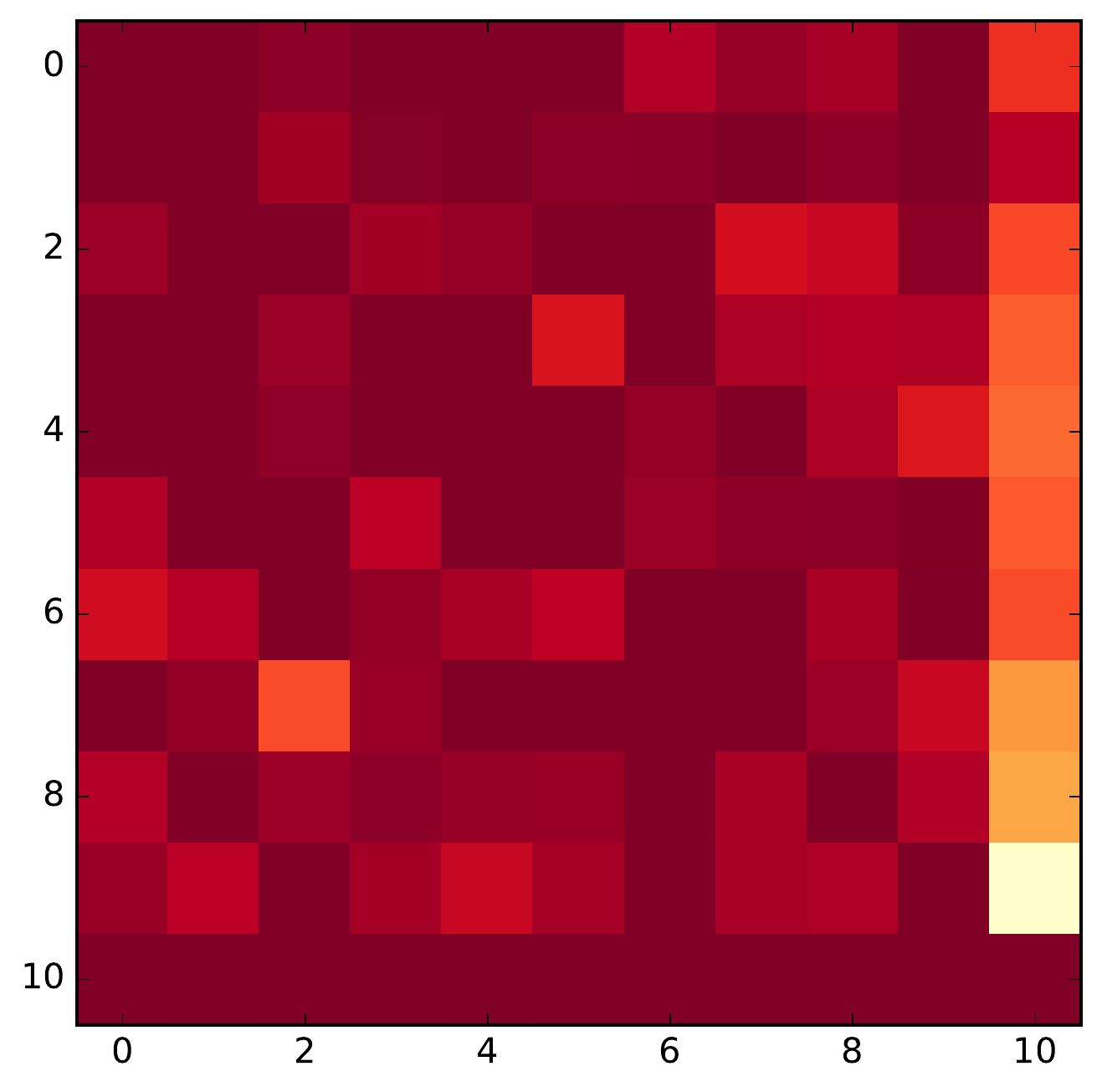}
\caption{Confusion matrices on benign test inputs of MNIST. The horizontal axis denotes the original label, and the vertical one the output class of the network. Left side corresponds to training on FGSM examples, right side on JSMA examples. The diagonal (correctly classified data points) has been zeroed, indices correspond to MNIST classes, where $10$ is the outlier class. Both matrices are normalized in the same scale, i.e., same color means same number of misclassified samples. Brighter indicates a higher number of misclassified examples.}
\label{fig:confmatrix}
\end{figure}

\boldpara{Wrongly classified benign test data} Next, we investigate the error
cases of our mechanism. The number of false positives, benign test examples of
the original data that are wrongly classified as outliers, represents a small
percentage of inputs: e.g., $0.5$\% on MNIST. In addition, we draw confusion
matrices for the benign test data in Figure~\ref{fig:confmatrix}. The diagonal
indicates correctly classified examples and is canceled out to better visualize
out-of-diagonal and misclassified inputs. Misclassification between classes is
very similar in the original case and when training with the FGSM. In the
interest of space, we thus omit the confusion matrix for original data. In
contrast, when training on JSMA examples, a large fraction of misclassified data
points is no longer misclassified as a legitimate class, but wrongly classified
as outliers.


\section{Preventing the Arms Race}\label{section:armsRace}
A key challenge in ML security lies in the fact that no defense guarantees
resilience to future attack designs. This contrasts with ML privacy where
differential privacy guarantees withstand all
hypothetical adversaries. Such an arms race may only be broken by mechanisms
that have been proven to be secure in an expressive security model,
such as the one of differential privacy.

While providing any formal guarantees for the methods proposed here is
intrinsically hard given the nature of optimization problems solved
by ML algorithms, we evaluate here their resilience to adaptive strategies. We first show
that the statistical test still performs well when presented with a mixture of
benign and malicious inputs. We then demonstrate the robustness of our models
augmented with the outlier class to powerful black-box strategies that have
evaded previous defenses.

\vspace*{-0.1in}

\subsection{Robustness of the Statistical Test}
Though we introduced the statistical not as a defense, but as a tool to investigate the distribution of adversarial examples, one might perform it on a batch before submitting inputs to the ML
model. A natural question is then whether an adversary aware of this defense
could evade it by constructing adversarial examples simultaneously misleading
the model and the statistical test. 

Theoretically, a statistical test cannot be misled because adversarial
examples necessarily deviate from the expected distribution (see Section~\ref{section:problemsetting}). Yet, this assumes that the defender is
capable of running statistical tests on sufficiently large sample sets of
inputs. In that case, it is guaranteed that the null hypothesis
would be rejected. However, this may not always be the case in practice
as such sampling may require a potentially infinite number of inputs.
As such, we measure the confidence of our statistical test as it is
presented with more realistic sample sets of inputs in an effort to
demonstrate its robustness.  

We consider two scenarios were the attacker adapted its strategy. First,
the attacker might hide small numbers of adversarial examples among a
large number of benign samples. In the following, we thus investigate the statistical test's 
performance in detecting adversarial inputs in the presence of legitimate inputs. Second, we consider sample sets of adversarial examples, where the
adversary executed  
more than one adversarial crafting algorithm, like for instance the FGSM
and the JSMA. 

\begin{figure}
	\centering
	\includegraphics[width=\linewidth]{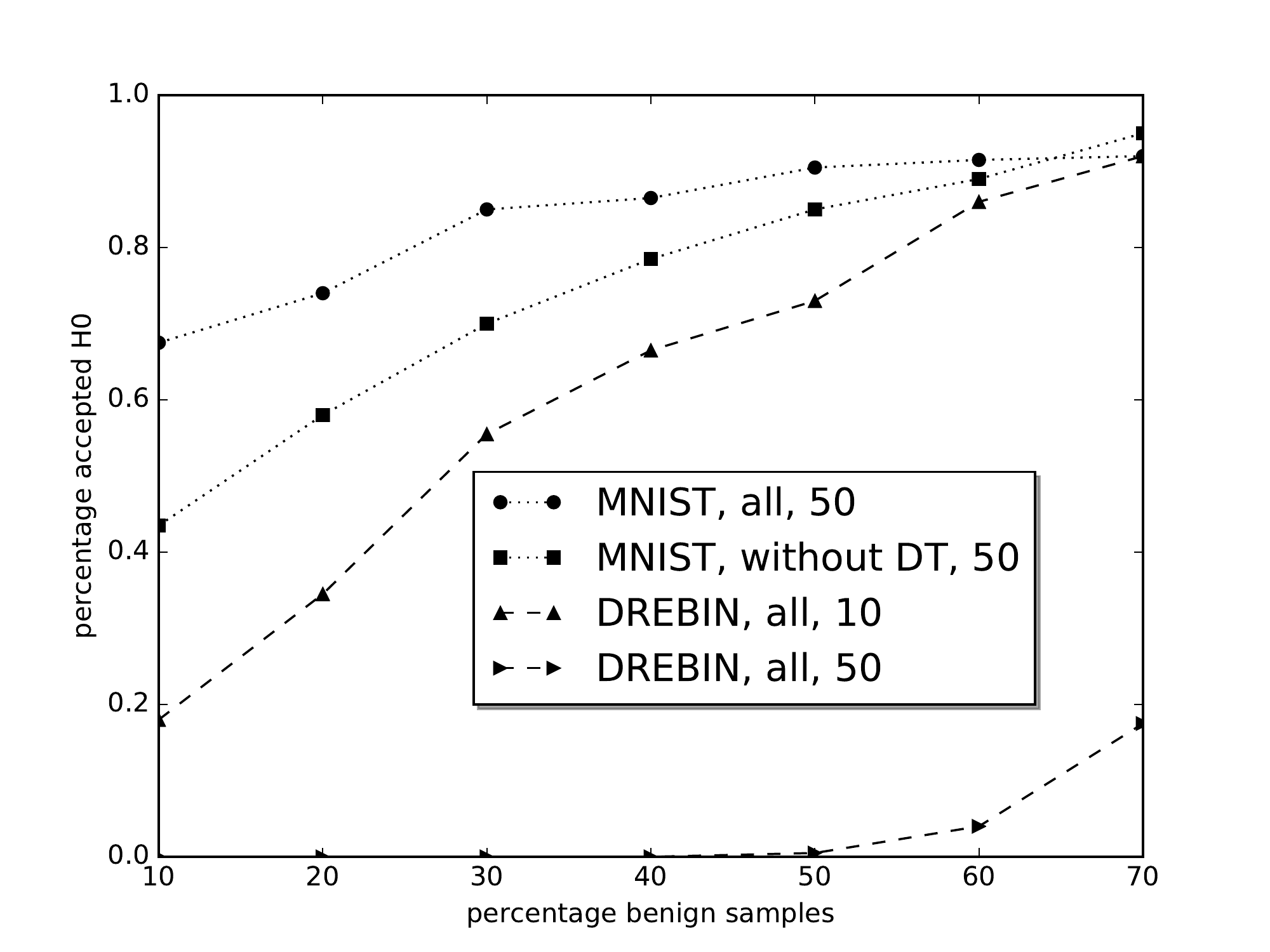}
	\caption{
	Evaluating mixtures of different adversarial examples and benign data. X-axis indicates the percentage of benign data. Y-axis is the percentage where the statistical test confirms that the data is from the same distribution (lower is better). We observe that the more benign examples (moving to the right), the harder it is to detect the remaining adversarial examples.}
	\label{fig:vanishing}
\end{figure}

\boldpara{Mixture of adversarial examples}We observe reductions in the
detection confidence for mixtures of adversarial examples. This is in particular the case when one of
the adversarial example kind is hard to detect. For instance, we
observed in Section~\ref{ssec:exp-adv-ex-test}  that adversarial inputs for
decision trees are hard to detect on MNIST. This reduces the performance of
the statistical test on samples that contain these examples. 
Our full results are depicted in Figure~\ref{fig:vanishing}. 

\boldpara{Mixtures of adversarial and legitimate inputs}The trade-off between the ratio of benign and adversarial examples, test confidence and sample size is shown in Figure~\ref{fig:vanishing}. The test is more confident when the percentage of adversarial examples is high or the sample size is large. Hence, one is less likely to detect adversarial examples mixed with legitimate inputs among small sets of inputs.

We conclude that the statistical test's confidence decreases, as expected,
when the adversary submits very few adversarial examples among large sets
of legitimate inputs. This is however
the main motivation behind the outlier class mechanism introduced in
Section~\ref{ssec:exp-improved-models}. 

\subsection{Robustness of the Outlier Class}
We now investigate the performance of models augmented with an outlier class
in the face of adversaries aware of that defense. We first show that these models are able to generalize to varying attacker strategies, i.e., they can detect adversarial inputs crafted using a different
algorithm than the one used to train the outlier class. 

In addition, we note that in all previous experiments, we considered adversaries directly computing adversarial
examples based on the defended model's parameters. Instead, we here evaluate
the model's robustness when attacked using black-box strategies. These
powerful attacks have been shown to evade previously proposed defenses, such as adversarial training and defensive distillation~\cite{papernot2016practical}. The reason is that these defenses
did not actually fix model errors but rather manipulated the model's gradients,
thus only making it harder for the adversary to craft adversarial examples
when they are computed directly on the targeted model. 

A simple but highly successful strategy is then for the adversary to train an auxiliary model that mimics the defended model's
predictions, and then use the auxiliary model to find adversarial examples
that are also misclassified by the defended model. In the following, we show
that our models with an outlier class are also robust to such strategies. 

\boldpara{Robustness of detection to adaptive attackers} 
We investigate whether the outlier class generalizes to
other adversarial example crafting techniques. In other words, we
ask whether defending against one type of adversarial examples is
sufficient to mitigate an \emph{adaptive} attacker using multiple techniques
to craft adversarial examples.

We thus trained the MNIST model's outlier class with only one kind of
adversarial example, and then observed its robustness to another kind of
adversarial examples. Table~\ref{table:ad} reports
this result for varying attack parameter intensities. If we train the model on JSMA adversarial examples, it is also robust to adversarial
FGSM examples crafted. If perturbations are high, misclassification is smaller than $3$\% percent. The reverse case, however, does not hold: a model trained on FGSM is only slightly more robust then the original model.  


We did not perform this experiment on DREBIN or MicroRNA datasets, since 
we could only apply one attack on each of them.

\begin{table}
\begin{tabular}{@{} lrrrrr@{} }
	\toprule[1.5pt]
\multicolumn{2}{c}{\emph{Training}} & \multicolumn{1}{c}{Attack} \\ \cmidrule(r){1-2} \cmidrule(r){3-3}
$\epsilon$ & Attack & $\epsilon$  & R & D & Error \\ \midrule
$\leq 200$ & JSMA & $0.1$  & $2.04$\% & $77.16$\% & $20.8$\% \\ 
$\leq 200$ & JSMA & $0.275$  & $2.07$\% & $96.6$\% & $2.95$\% \\
$ \leq 200$ & JSMA & $0.4$  & $0.22$\% & $98.45$\% & $1.33$\% \\
$ \leq 200$ & JSMA &$0.6$  & $0.13$\% & $99.58$\%  & $0.29$\% \\
$ 0.275$ & FGSM & $\leq 80$  & $0$\% & $9.63$\%  & $90.37$\%  \\ \bottomrule[1.5pt]
	\end{tabular}
	\caption{Misclassification and adaptive detection rate DNN trained with an outlier class on MNIST. All models are trained on $\epsilon = .275$ FGSM examples and $\epsilon < 200$ JSMA examples. The attacks used to evaluate the detection performance are different from the one used to train the outlier class. \emph{Recovered} (R) indicates the rate of adversarial
	examples classified in the original class of the input they were crafted from.
	\emph{Detected} (D) indicates the percentage of adversarial examples that were
	classified as outliers. The \emph{error} rate is simply the remaining
	adversarial examples (those not correctly classified or detected as outliers)} \label{table:ad}
\end{table}

\boldpara{Robustness of detection to black-box attacks performed using transferability}
We now show that our proposed outlier class
mechanism is robust to an additional attack vector against ML models: black-box attacks
exploiting adversarial example transferability. These techniques allow an
adversary to force a ML model to misclassify without knowledge of its model
parameters (and sometimes even without knowledge of its training data) by
computing adversarial examples on a different model than the one
targeted~\cite{szegedy2013intriguing,goodfellow2015explaining,papernot2016practical}.

In order to simulate the \emph{worst-case} adversary, we train the \emph{substitute} model
from which we will transfer adversarial examples back to the victim model
with the same training data than the one used by the victim. Again, a black-box attack may be possible even
without such knowledge~\cite{papernot2016practical}. Yet, this allows us
to consider a particularly strong adversary capable to closely mimic our model. 
We further focus on the robustness of our approach to such attacks, rather
than a demonstration of the feasibility of such attacks. 
We train two substitute models, one including the outlier class (denoted BB+1) as does the victim, the other not (BB). 
We then compute adversarial examples on the substitute model and evaluate the misclassification rates of the victim model.

\boldpara{MNIST results} The results on MNIST are given in Table~\ref{table:bbt}. We observe in both settings high robustness against the adversarial examples computed using the FSGM. Depending on the setting, we misclassify not a single of the examples when $\epsilon>0.4$ for BB+1 and $\epsilon>0.275$ for BB. We further observe high robustness as well concerning the adversarial examples computed using JSMA: In the setting of BB+1, the misclassification is around $6.7$\%, for BB even $2.7$\%. In all cases, most of the adversarial examples are classified as outliers. 


\begin{table}
	\begin{tabular}{@{}lrrrr @{} }
	\toprule[1.5pt]
    $\epsilon$ & Attack & R & D & E  \\ \midrule
    $0.1$ & FGSM & $42,67$\% & $55.72$\% &  $1.61$\%  \\
    $0.275$ & FGSM & $0$\% & $100$\%  & $0$\% \\
    $> 0.4$ & FGSM &$0$\% & $100$\% & $0$\%  \\
    $ \leq 80$ & JSMA & $0.3$\% & $97$\%  & $2.7$\%  \\ \bottomrule[1.5pt]   
	\end{tabular}
	\vspace*{1ex}
	\begin{tabular}{@{}lrrrr @{} }
	\toprule[1.5pt]
    $\epsilon$ & Attack & R & D & E  \\ \midrule
    $0.1$  & FGSM & $17.64$\% & $81.64$\% & $0.72$\% \\
    $0.275$ & FGSM & $0$\% & $99.98$\%  & $0.02$\%\\
    $> 0.4$ & FGSM &$0$\% & $100$\% & $0$\% \\
    $ \leq 80$& JSMA & $0.46$\% & $92.85$\% & $6.69$\%  \\ \bottomrule[1.5pt]   
	\end{tabular}
			\caption{Robustness of a MNIST model (with an outlier class) to black-box attackers. All models are trained on $\epsilon = .275$ FGSM examples and $\epsilon < 200$ JSMA examples. \emph{Recovered} (R) indicates the     rate of adversarial
	examples classified in the original class of the input they were crafted from.
	\emph{Detected} (D) indicates the percentage of adversarial examples that were
	classified as outliers. The \emph{Error} (E) rate is simply the remaining
	adversarial examples (those not correctly classified or detected as outliers). Above Table show an attacker using a substitute model without an outlier class. Lower one indicates an attacker using a substitute model with an outlier class, equivalent to tested model.}
    \label{table:bbt}

\end{table}

\begin{table*}[t]
	\centering{\small
	\begin{tabular}{@{}lrrrrrrrrrrrr @{}}
		\toprule[1.5pt]
		& \multicolumn{6}{c}{BB} & \multicolumn{6}{c}{BB+1} \\ \cmidrule(r){2-7} \cmidrule(l){8-13}
	    & \multicolumn{3}{c}{logistic regression} & \multicolumn{3}{c}{neural network} & \multicolumn{3}{c}{logistic regression} & \multicolumn{3}{c}{neural network} \\ 
		\cmidrule(r){2-4} \cmidrule(r){5-7} \cmidrule(r){8-10} \cmidrule(l){11-13}
		$\epsilon$ & R & D & E & R & D & E & R & D & E & R & D & E\\
	    \midrule
		$0.2$ & $84.6$\% & $11.9$\% & $3.5$\% & $83.6$\% & $12.9$\% & $3.5$\% & $81.3$\% & $14.9$\% & $3.8$\% & $80.8$\% & $15.7$\% & $3.5$\% \\
		$0.4$ & $49,7$\% &  $46.7$\% & $3.6$\% & $47.7$\% & $48.7$\% & $3.6$\% & $35.1$\% & $59.8$\% & $5.1$\% & $27.8$\% & $65.2$\% & $5$\%\\ 
		$0.6$ & $21.5$\% & $74.7$\% & $3.8$\% & $15.4$\% & $80.6$\% &$4$\% & $9$\% & $74.5$\% &$16.5$\% & $4.5$\% & $68.9$\% & $26.6$\% \\  
		$0.8$  & $14.8$\% & $75.3$\% & $9.8$\% & $7.3$\% & $81.8$\% & $10.9$\% & $6$\% & $55.8$\% & $38.2$\% & $4$\%& $30.8$\% & $65.2$\% \\ 
		$1.0$ & $9.3$\% & $68.2$\%  & $22.5$\% & $3.8$\% & $65.4$\%&$30.8$\% & $4.3$\% & $38.6$\%  & $57.1$\% & $6$\% & $6.0$\% & $88$\% \\	
		\midrule
		$0.2$ & $83.3$\% & $13.9$\% &  $2.8$\% & $82.6$\% & $14.6$\% &  $2.8$\% & $76.3$\% & $21.5$\% & $2.2$\% & $84.1$\% & $12.1$\% &$3.8$\% \\
		$0.4$ & $45.2$\% & $54.0$\% & $1.8$\% & $42.9$\% & $55.3$\% & $1.8$\% & $30.3$\% & $68.4$\% & $1.3$\% &$50.3$\% & $45.7$\%& $4$\% \\ 
		$0.6$ & $23.4$\% & $76.0$\%&$10.6$\% & $16.5$\% & $78.3$ \% & $5$\% & $4$\% & $93.2$\% & $2.8$\% & $14.6$\% & $81.6$\% &$3.8$\%\\  
		$0.8$  & $13.4$\% & $76$\%& $10.6$\% & $6$\% & $80.3$\% &$13.7$\% & $2$\% & $94.2$\% & $3.8$\% & $3.2$\% & $95.0$\% & $1.8$\% \\ 
		$1.0$ & $3$\% & $79.8$\% & $17.2$\% & $2$\% & $73.2$\% & $24.8$\% & $2$\% &$94.7$\% &$3.3$\% & $2$\% & $97.2$\% & $0.8$\% \\	\bottomrule[1.5pt]
	\end{tabular}}
	\vspace*{2ex}
	\caption{Black box setting for logistic regression (upper part) and a neural network (lower part) on the MicroRNA data trained using the outlier class. Substitutes are logistic regression or a neural network (NN), either trained without (BB) or with an additional class (BB+1). If an outlier class is used,  FGSM examples at $\epsilon=1.0$ and $\epsilon=0.8$ are used for training. We report parameters of attack ($\epsilon$). \emph{Recovered} (R) indicates the     rate of adversarial
	examples classified in the original class of the input they were crafted from.
	\emph{Detected} (D) indicates the percentage of adversarial examples that were
	classified as outliers. The \emph{Error}(E) rate is simply the remaining
	adversarial examples (those not correctly classified or detected as outliers) 
	.}\label{table:mircorna_bb}
\end{table*}

\boldpara{DREBIN results} 
We observe that, though the network is vulnerable to direct attacks, it is much more resistant to adversarial examples crafted on another network. In this setting, we only compute JSMA examples. Given a substitute trained with additional class,
we observe misclassification rates of $35.8$\%, $47.1$\% are detected as outliers. For JSMA examples crafted on the original network $0.8$\% are misclassified; $98.6$\% of adversarial examples are classified as outliers.

\boldpara{MicroRNA results} The results on the MicroRNA data are depicted in Table~\ref{table:mircorna_bb}. We observe that logistic regression works well in detecting smaller perturbations till $0.4$. For larger $\epsilon$, however, the misclassification drastically increases, partially to $88$\%. Again, we assign the small improvement to the limited capacity of logistic regression, we trained additionally neural networks. We observed lower misclassification in 
settings where the perturbation is maximmal ($\epsilon=1.0$). In general, we observe that logistic regression is more robust in the BB setting, whereas neural networks are more robust in the BB+1 setting (both with some exceptions).

\section{Discussion}\label{section:discussion}
We discuss the limitations of the  mechanisms proposed to detect adversarial
examples: statistical testing, and an outlier class. We also explore avenues for
future work.

\boldpara{Statistical Test} As we have seen, one of the major strengths of
kernel-based statistical tests is that they operate and thus detect the presence
of adversarial examples already in feature space, before these inputs are even
fed to the ML model. Intuitively, we observed that the larger the perturbation
applied is, the more likely it is to be confidently detected by the statistical
test. Adversarial example crafting techniques that modify few features (like the
JSMA or the decision tree attack) or perturb the features only slightly (small
values of $\epsilon$ for the FGSM) are less likely to be detected.

This finding is consistent with the underlying stationary assumption made by all
ML approaches. Since adversarial examples are not drawn from the same
distribution than benign data, the classifier is incapable of classifying them
correctly. This property also holds for the training data itself, and as such,
we expect it to generalize to poisoning attacks. In
such attacks, the adversary attempts to degrade learning by inserting malicious
points in the model's training data. This is however outside the scope of this
work, and we leave this question to future work.

\boldpara{Integrating Outlier Detection}
We further observe that adding an outlier class to the model yields robustness to adaptive attack strategies, and needed perturbation is increased. Concerning JSMA, for some datasets, we do not achieve robustness to adversarial examples. This most likely depends on the initial vulnerability of the data: For DREBIN we changed barely more than one feature, for MNIST almost twenty. At the same time, MNIST has slightly less features, of which the pixels at the borders are barely used. Thus, having less features and a higher perturbation to learn from might yield larger robustness to adversarial examples. Additionally, further factors might include inter-class and intra-class distances, or the variability of the computed adversarial examples.

Further, the confusion matrices from Figure~\ref{fig:confmatrix} seem to suggest that FGSM lie in a different halfspace than the original data: the outlier class trained on JSMA examples contains benign data points whereas the FGSM one does not. This might indicate that JSMA examples lie rather between benign classes.

We further observed that knowledge about the attack is not necessarily needed: training on adversarial examples computed using the JSMA hardens against computing FGSM examples. Perhaps surprisingly, this does not hold the other way around. In general, since FGSM is non-targeted and less optimal than JSMA, further work is needed whether the outlier class generalizes from targeted to non-targeted attacks or from more optimal to less optimal attacks. In theory, we could feed samples from the whole feature space except the location of the classes, and thus obtain a robust classifier without any assumption on the adversary. This is practically infeasible, however. Future work will investigate trade-offs here.

Finally, we want to remark that the benign data, that is labeled as outlier by the network might be beneficial when investigated by an expert\cite{DBLP:conf/dimva/MillerKTABFHSWY16}. This data might be either excluded from training, or relabeled. This question will be answered in future work.

\section{Related Work}
\label{section:related-work}

Other approaches to detect malicious data points by
statistical means have been proposed. However, they all depend on some of the
internal activations of deep neural networks
models~\cite{DBLP:journals/corr/LiL16e,DBLP:conf/acsac/ShenTS16,2017FeinmanDASfA}. Hence, these
approaches only apply to the specific classifier studied. In contrast, we apply
our statistical test directly in feature space, allowing us to propose a
model-agnostic detection.

Wang et al.~\cite{DBLP:journals/corr/WangGQ16} present a similar formal intuition as we do (using an oracle instead of the underlying distribution, though). From this, they formally derive conditions when a classifier is secured against adversarial examples. Further, they proposed a modified version of adversarial training, originally introduced by Goodfellow et al.~\cite{goodfellow2015explaining}. In contrast to both of these approaches, we classify adversarial examples in a separate (and additional) outlier class. We also do not compute adversarial examples throughout training but rather use adversarial examples precomputed on a different model before training.

Nguyen et al \cite{DBLP:conf/cvpr/NguyenYC15} have introduced an outlier class before. Also Bendale et al \cite{DBLP:conf/cvpr/BendaleB16} propse open networks, that are not confident in their classification all over the feature space. Both, however, do not evaluate and motivate their approach in adversarial settings.

Metzen et al.~\cite{addClassifier} augment neural networks with 
an auxiliary network used to detect malicious samples. This additional network shares some of its parameters with the original one, and thus also depends on the features of the network. Further,  our outlier class mechanism is applicable to any ML models. In addition, their approach
is limited in settings where the adversary adapts its strategy.
Instead, our experiments systematically explore the space of adversaries (with
different adversarial example crafting algorithms, datasets and models).
We also present a detailed discussion of possible adaptive strategies, such as 
powerful black-box attacks known to be hard to defend against~\cite{papernot2016practical}. 


\vspace*{-0.1in}

\section{Conclusion}\label{section:conclusion}
\vspace*{-0.05in}
We empirically validated the hypothesis that adversarial examples can be
detected using statistical tests before they are even fed to the ML model as
inputs. Thus, their malicious properties are model-agnostic. 

Furthermore, we show how to augment ML models with an additional class in which
the model is trained to classify all adversarial inputs. This results in
robustness to adversaries, even those using attack strategies based on
transferability---a class of attacks known to be harder to defend against than
gradient-based strategies. In addition, when adversarial examples with small 
perturbations are not detected as outliers, they are original class is often recovered and the perturbed input correctly classified.

Additionally, we expect that combining our approaches together, as well as with
other defenses may prove beneficial. For instance, we expect defensive
distillation and the statistical test or outlier class to work well together, as
defensive distillation has been found to increase the perturbations that an
adversary introduces.

\section*{Acknowledgments}

Nicolas Papernot is supported
by a Google PhD Fellowship in Security.
Research was supported in part by the Army Research Laboratory,
under Cooperative Agreement Number W911NF-13-2-0045 (ARL Cyber Security
CRA), and the Army Research Office under grant W911NF-13-1-0421.
The views and conclusions contained in this document are those of the
authors and should not be interpreted as representing the official policies,
either expressed or implied, of the Army Research Laboratory or the U.S.
Government. The U.S.\ Government is authorized to reproduce and distribute
reprints for government purposes notwithstanding any copyright notation hereon.

This work was supported by the German Federal Ministry of Education and
Research (BMBF) through funding for the Center for IT-Security,
Privacy and Accountability (CISPA) (FKZ: 16KIS0753). 

\bibliographystyle{ACM-Reference-Format}
\bibliography{lit}


\begin{thebibliography}{00}


\ifx \showCODEN    \undefined \def \showCODEN     #1{\unskip}     \fi
\ifx \showDOI      \undefined \def \showDOI       #1{#1}\fi
\ifx \showISBNx    \undefined \def \showISBNx     #1{\unskip}     \fi
\ifx \showISBNxiii \undefined \def \showISBNxiii  #1{\unskip}     \fi
\ifx \showISSN     \undefined \def \showISSN      #1{\unskip}     \fi
\ifx \showLCCN     \undefined \def \showLCCN      #1{\unskip}     \fi
\ifx \shownote     \undefined \def \shownote      #1{#1}          \fi
\ifx \showarticletitle \undefined \def \showarticletitle #1{#1}   \fi
\ifx \showURL      \undefined \def \showURL       {\relax}        \fi
\providecommand\bibfield[2]{#2}
\providecommand\bibinfo[2]{#2}
\providecommand\natexlab[1]{#1}
\providecommand\showeprint[2][]{arXiv:#2}

\bibitem[\protect\citeauthoryear{A.~shimomura}{A.~shimomura}{2016}]%
        {RNAData}
\bibfield{author}{\bibinfo{person}{Kawauchi J Takizawa S et~al A.~shimomura,
  Shiino~S}.} \bibinfo{year}{2016}\natexlab{}.
\newblock \showarticletitle{Novel combination of serum microRNA for detecting
  breast cancer in the early stage}.
\newblock \bibinfo{journal}{{\em Cancer Sci Mar\/}}
  \bibinfo{volume}{107(3):326-34} (\bibinfo{year}{2016}).
\newblock


\bibitem[\protect\citeauthoryear{Arp, Spreitzenbarth, Hubner, Gascon, and
  Rieck}{Arp et~al\mbox{.}}{2014}]%
        {arp2014drebin}
\bibfield{author}{\bibinfo{person}{Daniel Arp}, \bibinfo{person}{Michael
  Spreitzenbarth}, \bibinfo{person}{Malte Hubner}, \bibinfo{person}{Hugo
  Gascon}, {and} \bibinfo{person}{Konrad Rieck}.}
  \bibinfo{year}{2014}\natexlab{}.
\newblock \showarticletitle{{DREBIN: Effective and Explainable Detection of
  Android Malware in Your Pocket.}}. In \bibinfo{booktitle}{{\em Proceedings of
  the 2014 Network and Distributed System Security Symposium (NDSS)}}.
\newblock


\bibitem[\protect\citeauthoryear{Bendale and Boult}{Bendale and Boult}{2016}]%
        {DBLP:conf/cvpr/BendaleB16}
\bibfield{author}{\bibinfo{person}{Abhijit Bendale} {and}
  \bibinfo{person}{Terrance~E. Boult}.} \bibinfo{year}{2016}\natexlab{}.
\newblock \showarticletitle{Towards Open Set Deep Networks}. In
  \bibinfo{booktitle}{{\em 2016 {IEEE} Conference on Computer Vision and
  Pattern Recognition, {CVPR} 2016, Las Vegas, NV, USA, June 27-30, 2016}}.
  \bibinfo{pages}{1563--1572}.
\newblock
\showDOI{%
\url{https://doi.org/10.1109/CVPR.2016.173}}


\bibitem[\protect\citeauthoryear{Biggio, Nelson, and Laskov}{Biggio
  et~al\mbox{.}}{2012}]%
        {biggio2012poisoning}
\bibfield{author}{\bibinfo{person}{Battista Biggio}, \bibinfo{person}{Blaine
  Nelson}, {and} \bibinfo{person}{Pavel Laskov}.}
  \bibinfo{year}{2012}\natexlab{}.
\newblock \showarticletitle{Poisoning attacks against support vector machines}.
\newblock \bibinfo{journal}{{\em arXiv preprint arXiv:1206.6389\/}}
  (\bibinfo{year}{2012}).
\newblock


\bibitem[\protect\citeauthoryear{Bousquet and Elisseeff}{Bousquet and
  Elisseeff}{2002}]%
        {Bousquet02}
\bibfield{author}{\bibinfo{person}{Olivier Bousquet} {and}
  \bibinfo{person}{Andr{\'e} Elisseeff}.} \bibinfo{year}{2002}\natexlab{}.
\newblock \showarticletitle{Stability and Generalization}.
\newblock \bibinfo{journal}{{\em The Journal of Machine Learning Research\/}}
  \bibinfo{volume}{2} (\bibinfo{year}{2002}), \bibinfo{pages}{499--526}.
\newblock


\bibitem[\protect\citeauthoryear{Br{\"{u}}ckner and Scheffer}{Br{\"{u}}ckner
  and Scheffer}{2011}]%
        {DBLP:conf/kdd/BrucknerS11}
\bibfield{author}{\bibinfo{person}{Michael Br{\"{u}}ckner} {and}
  \bibinfo{person}{Tobias Scheffer}.} \bibinfo{year}{2011}\natexlab{}.
\newblock \showarticletitle{Stackelberg games for adversarial prediction
  problems}. In \bibinfo{booktitle}{{\em Proceedings of the 17th {ACM} {SIGKDD}
  International Conference on Knowledge Discovery and Data Mining, San Diego,
  CA, USA, August 21-24, 2011}}. \bibinfo{pages}{547--555}.
\newblock
\showDOI{%
\url{https://doi.org/10.1145/2020408.2020495}}


\bibitem[\protect\citeauthoryear{Carlini and Wagner}{Carlini and
  Wagner}{2016}]%
        {DBLP:journals/corr/CarliniW16a}
\bibfield{author}{\bibinfo{person}{Nicholas Carlini} {and}
  \bibinfo{person}{David Wagner}.} \bibinfo{year}{2016}\natexlab{}.
\newblock \showarticletitle{Towards Evaluating the Robustness of Neural
  Networks}.
\newblock \bibinfo{journal}{{\em CoRR\/}}  \bibinfo{volume}{abs/1608.04644}
  (\bibinfo{year}{2016}).
\newblock
\showURL{%
\url{http://arxiv.org/abs/1608.04644}}


\bibitem[\protect\citeauthoryear{{Carlini} and {Wagner}}{{Carlini} and
  {Wagner}}{2017}]%
        {2017arXiv170507263C}
\bibfield{author}{\bibinfo{person}{N. {Carlini}} {and} \bibinfo{person}{D.
  {Wagner}}.} \bibinfo{year}{2017}\natexlab{}.
\newblock \showarticletitle{{Adversarial Examples Are Not Easily Detected:
  Bypassing Ten Detection Methods}}.
\newblock \bibinfo{journal}{{\em ArXiv e-prints\/}} (\bibinfo{date}{May}
  \bibinfo{year}{2017}).
\newblock
\showeprint[arxiv]{cs.LG/1705.07263}


\bibitem[\protect\citeauthoryear{Dalvi, Domingos, Sanghai, Verma,
  et~al\mbox{.}}{Dalvi et~al\mbox{.}}{2004}]%
        {dalvi2004adversarial}
\bibfield{author}{\bibinfo{person}{Nilesh Dalvi}, \bibinfo{person}{Pedro
  Domingos}, \bibinfo{person}{Sumit Sanghai}, \bibinfo{person}{Deepak Verma},
  {et~al\mbox{.}}} \bibinfo{year}{2004}\natexlab{}.
\newblock \showarticletitle{Adversarial classification}. In
  \bibinfo{booktitle}{{\em Proceedings of the tenth ACM SIGKDD international
  conference on Knowledge discovery and data mining}}. ACM,
  \bibinfo{pages}{99--108}.
\newblock


\bibitem[\protect\citeauthoryear{Friedman and Rafsky}{Friedman and
  Rafsky}{1979}]%
        {Friedman_Rafsky_1979}
\bibfield{author}{\bibinfo{person}{J.H. Friedman} {and} \bibinfo{person}{L.C.
  Rafsky}.} \bibinfo{year}{1979}\natexlab{}.
\newblock \showarticletitle{Multivariate generalizations of the Wald--Wolfowitz
  and Smirnov two-sample tests}.
\newblock \bibinfo{journal}{{\em Ann. Stat.; (United States)\/}}
  \bibinfo{volume}{7:4} (\bibinfo{date}{Jan} \bibinfo{year}{1979}).
\newblock


\bibitem[\protect\citeauthoryear{Goodfellow et~al\mbox{.}}{Goodfellow
  et~al\mbox{.}}{2015}]%
        {goodfellow2015explaining}
\bibfield{author}{\bibinfo{person}{Ian~J Goodfellow} {et~al\mbox{.}}}
  \bibinfo{year}{2015}\natexlab{}.
\newblock \showarticletitle{Explaining and Harnessing Adversarial Examples}. In
  \bibinfo{booktitle}{{\em Proceedings of the 2015 International Conference on
  Learning Representations}}.
\newblock


\bibitem[\protect\citeauthoryear{Goodfellow, Papernot, and McDaniel}{Goodfellow
  et~al\mbox{.}}{2016}]%
        {DBLP:journals/corr/GoodfellowPM16}
\bibfield{author}{\bibinfo{person}{Ian~J. Goodfellow}, \bibinfo{person}{Nicolas
  Papernot}, {and} \bibinfo{person}{Patrick~D. McDaniel}.}
  \bibinfo{year}{2016}\natexlab{}.
\newblock \showarticletitle{cleverhans v0.1: an adversarial machine learning
  library}.
\newblock \bibinfo{journal}{{\em CoRR\/}}  \bibinfo{volume}{abs/1610.00768}
  (\bibinfo{year}{2016}).
\newblock
\showURL{%
\url{http://arxiv.org/abs/1610.00768}}


\bibitem[\protect\citeauthoryear{Gretton, Borgwardt, Rasch, Sch{\"{o}}lkopf,
  and Smola}{Gretton et~al\mbox{.}}{2012}]%
        {DBLP:journals/jmlr/GrettonBRSS12}
\bibfield{author}{\bibinfo{person}{Arthur Gretton}, \bibinfo{person}{Karsten~M.
  Borgwardt}, \bibinfo{person}{Malte~J. Rasch}, \bibinfo{person}{Bernhard
  Sch{\"{o}}lkopf}, {and} \bibinfo{person}{Alexander~J. Smola}.}
  \bibinfo{year}{2012}\natexlab{}.
\newblock \showarticletitle{A Kernel Two-Sample Test}.
\newblock \bibinfo{journal}{{\em Journal of Machine Learning Research\/}}
  \bibinfo{volume}{13} (\bibinfo{year}{2012}), \bibinfo{pages}{723--773}.
\newblock
\showURL{%
\url{http://dl.acm.org/citation.cfm?id=2188410}}


\bibitem[\protect\citeauthoryear{Grosse, Papernot, Manoharan, Backes, and
  McDaniel}{Grosse et~al\mbox{.}}{2016}]%
        {DBLP:journals/corr/GrossePM0M16}
\bibfield{author}{\bibinfo{person}{Kathrin Grosse}, \bibinfo{person}{Nicolas
  Papernot}, \bibinfo{person}{Praveen Manoharan}, \bibinfo{person}{Michael
  Backes}, {and} \bibinfo{person}{Patrick McDaniel}.}
  \bibinfo{year}{2016}\natexlab{}.
\newblock \showarticletitle{Adversarial Perturbations Against Deep Neural
  Networks for Malware Classification}.
\newblock \bibinfo{journal}{{\em CoRR\/}}  \bibinfo{volume}{abs/1606.04435}
  (\bibinfo{year}{2016}).
\newblock
\showURL{%
\url{http://arxiv.org/abs/1606.04435}}


\bibitem[\protect\citeauthoryear{Hall and Tajvidi}{Hall and Tajvidi}{2002}]%
        {hall2002permutation}
\bibfield{author}{\bibinfo{person}{Peter Hall} {and} \bibinfo{person}{Nader
  Tajvidi}.} \bibinfo{year}{2002}\natexlab{}.
\newblock \showarticletitle{Permutation tests for equality of distributions in
  high-dimensional settings}.
\newblock \bibinfo{journal}{{\em Biometrika\/}} \bibinfo{volume}{89},
  \bibinfo{number}{2} (\bibinfo{year}{2002}), \bibinfo{pages}{359--374}.
\newblock


\bibitem[\protect\citeauthoryear{Hotelling}{Hotelling}{1931}]%
        {hotelling1931}
\bibfield{author}{\bibinfo{person}{Harold Hotelling}.}
  \bibinfo{year}{1931}\natexlab{}.
\newblock \showarticletitle{The Generalization of Student's Ratio}.
\newblock \bibinfo{journal}{{\em Ann. Math. Statist.\/}} \bibinfo{volume}{2},
  \bibinfo{number}{3} (\bibinfo{date}{08} \bibinfo{year}{1931}),
  \bibinfo{pages}{360--378}.
\newblock
\showDOI{%
\url{https://doi.org/10.1214/aoms/1177732979}}


\bibitem[\protect\citeauthoryear{Huang, Joseph, Nelson, Rubinstein, and
  Tygar}{Huang et~al\mbox{.}}{2011}]%
        {DBLP:conf/ccs/HuangJNRT11}
\bibfield{author}{\bibinfo{person}{Ling Huang}, \bibinfo{person}{Anthony~D.
  Joseph}, \bibinfo{person}{Blaine Nelson}, \bibinfo{person}{Benjamin I.~P.
  Rubinstein}, {and} \bibinfo{person}{J.~D. Tygar}.}
  \bibinfo{year}{2011}\natexlab{}.
\newblock \showarticletitle{Adversarial machine learning}. In
  \bibinfo{booktitle}{{\em Proceedings of the 4th {ACM} Workshop on Security
  and Artificial Intelligence, AISec 2011, Chicago, IL, USA, October 21,
  2011}}. \bibinfo{pages}{43--58}.
\newblock
\showDOI{%
\url{https://doi.org/10.1145/2046684.2046692}}


\bibitem[\protect\citeauthoryear{Jan Hendrick~Metzen and Bischoff}{Jan
  Hendrick~Metzen and Bischoff}{2016}]%
        {addClassifier}
\bibfield{author}{\bibinfo{person}{Volker~Fischer Jan Hendrick~Metzen,
  Tim~Genewein} {and} \bibinfo{person}{Bastian Bischoff}.}
  \bibinfo{year}{2016}\natexlab{}.
\newblock \showarticletitle{On detecting Adversarial Perturbations}.
\newblock  (\bibinfo{year}{2016}).
\newblock
\newblock
\shownote{(to appear).}


\bibitem[\protect\citeauthoryear{LeCun, Bottou, Bengio, and Haffner}{LeCun
  et~al\mbox{.}}{1998}]%
        {lecun-98}
\bibfield{author}{\bibinfo{person}{Y. LeCun}, \bibinfo{person}{L. Bottou},
  \bibinfo{person}{Y. Bengio}, {and} \bibinfo{person}{P. Haffner}.}
  \bibinfo{year}{1998}\natexlab{}.
\newblock \showarticletitle{Gradient-Based Learning Applied to Document
  Recognition}.
\newblock \bibinfo{journal}{{\it Proc. IEEE}} \bibinfo{volume}{86},
  \bibinfo{number}{11} (\bibinfo{date}{November} \bibinfo{year}{1998}),
  \bibinfo{pages}{2278--2324}.
\newblock


\bibitem[\protect\citeauthoryear{Li and Li}{Li and Li}{2016}]%
        {DBLP:journals/corr/LiL16e}
\bibfield{author}{\bibinfo{person}{Xin Li} {and} \bibinfo{person}{Fuxin Li}.}
  \bibinfo{year}{2016}\natexlab{}.
\newblock \showarticletitle{Adversarial Examples Detection in Deep Networks
  with Convolutional Filter Statistics}.
\newblock \bibinfo{journal}{{\em CoRR\/}}  \bibinfo{volume}{abs/1612.07767}
  (\bibinfo{year}{2016}).
\newblock
\showURL{%
\url{http://arxiv.org/abs/1612.07767}}


\bibitem[\protect\citeauthoryear{Liu and Chawla}{Liu and Chawla}{2010}]%
        {DBLP:journals/ml/LiuC10}
\bibfield{author}{\bibinfo{person}{Wei Liu} {and} \bibinfo{person}{Sanjay
  Chawla}.} \bibinfo{year}{2010}\natexlab{}.
\newblock \showarticletitle{Mining adversarial patterns via regularized loss
  minimization}.
\newblock \bibinfo{journal}{{\em Machine Learning\/}} \bibinfo{volume}{81},
  \bibinfo{number}{1} (\bibinfo{year}{2010}), \bibinfo{pages}{69--83}.
\newblock
\showDOI{%
\url{https://doi.org/10.1007/s10994-010-5199-2}}


\bibitem[\protect\citeauthoryear{Lowd and Meek}{Lowd and Meek}{2005}]%
        {DBLP:conf/ceas/LowdM05}
\bibfield{author}{\bibinfo{person}{Daniel Lowd} {and}
  \bibinfo{person}{Christopher Meek}.} \bibinfo{year}{2005}\natexlab{}.
\newblock \showarticletitle{Good Word Attacks on Statistical Spam Filters}. In
  \bibinfo{booktitle}{{\em {CEAS} 2005 - Second Conference on Email and
  Anti-Spam, July 21-22, 2005, Stanford University, California, {USA}}}.
\newblock
\showURL{%
\url{http://www.ceas.cc/papers-2005/125.pdf}}


\bibitem[\protect\citeauthoryear{Miller, Kantchelian, Tschantz, Afroz,
  Bachwani, Faizullabhoy, Huang, Shankar, Wu, Yiu, Joseph, and Tygar}{Miller
  et~al\mbox{.}}{2016}]%
        {DBLP:conf/dimva/MillerKTABFHSWY16}
\bibfield{author}{\bibinfo{person}{Brad Miller}, \bibinfo{person}{Alex
  Kantchelian}, \bibinfo{person}{Michael~Carl Tschantz}, \bibinfo{person}{Sadia
  Afroz}, \bibinfo{person}{Rekha Bachwani}, \bibinfo{person}{Riyaz
  Faizullabhoy}, \bibinfo{person}{Ling Huang}, \bibinfo{person}{Vaishaal
  Shankar}, \bibinfo{person}{Tony Wu}, \bibinfo{person}{George Yiu},
  \bibinfo{person}{Anthony~D. Joseph}, {and} \bibinfo{person}{J.~D. Tygar}.}
  \bibinfo{year}{2016}\natexlab{}.
\newblock \showarticletitle{Reviewer Integration and Performance Measurement
  for Malware Detection}. In \bibinfo{booktitle}{{\em Detection of Intrusions
  and Malware, and Vulnerability Assessment - 13th International Conference,
  {DIMVA} 2016, San Sebasti{\'{a}}n, Spain, July 7-8, 2016, Proceedings}}.
  \bibinfo{pages}{122--141}.
\newblock
\showDOI{%
\url{https://doi.org/10.1007/978-3-319-40667-1_7}}


\bibitem[\protect\citeauthoryear{Nguyen, Yosinski, and Clune}{Nguyen
  et~al\mbox{.}}{2015}]%
        {DBLP:conf/cvpr/NguyenYC15}
\bibfield{author}{\bibinfo{person}{Anh~Mai Nguyen}, \bibinfo{person}{Jason
  Yosinski}, {and} \bibinfo{person}{Jeff Clune}.}
  \bibinfo{year}{2015}\natexlab{}.
\newblock \showarticletitle{Deep neural networks are easily fooled: High
  confidence predictions for unrecognizable images}. In
  \bibinfo{booktitle}{{\em {IEEE} Conference on Computer Vision and Pattern
  Recognition, {CVPR} 2015, Boston, MA, USA, June 7-12, 2015}}.
  \bibinfo{pages}{427--436}.
\newblock
\showDOI{%
\url{https://doi.org/10.1109/CVPR.2015.7298640}}


\bibitem[\protect\citeauthoryear{Papernot, McDaniel, Goodfellow, Jha, and
  al.}{Papernot et~al\mbox{.}}{2016b}]%
        {papernot2016practical}
\bibfield{author}{\bibinfo{person}{Nicolas Papernot}, \bibinfo{person}{Patrick
  McDaniel}, \bibinfo{person}{Ian Goodfellow}, \bibinfo{person}{Somesh Jha},
  {and} \bibinfo{person}{al.}} \bibinfo{year}{2016}\natexlab{b}.
\newblock \showarticletitle{Practical Black-Box Attacks against Deep Learning
  Systems using Adversarial Examples}.
\newblock \bibinfo{journal}{{\em arXiv preprint arXiv:1602.02697\/}}
  (\bibinfo{year}{2016}).
\newblock


\bibitem[\protect\citeauthoryear{Papernot, McDaniel, and Goodfellow}{Papernot
  et~al\mbox{.}}{2016a}]%
        {DBLP:journals/corr/PapernotMG16}
\bibfield{author}{\bibinfo{person}{Nicolas Papernot}, \bibinfo{person}{Patrick
  McDaniel}, {and} \bibinfo{person}{Ian~J. Goodfellow}.}
  \bibinfo{year}{2016}\natexlab{a}.
\newblock \showarticletitle{Transferability in Machine Learning: from Phenomena
  to Black-Box Attacks using Adversarial Samples}.
\newblock \bibinfo{journal}{{\em CoRR\/}}  \bibinfo{volume}{abs/1605.07277}
  (\bibinfo{year}{2016}).
\newblock
\showURL{%
\url{http://arxiv.org/abs/1605.07277}}


\bibitem[\protect\citeauthoryear{Papernot, McDaniel, Jha, Fredrikson, Celik,
  and Swami}{Papernot et~al\mbox{.}}{2016c}]%
        {papernot2016limitations}
\bibfield{author}{\bibinfo{person}{Nicolas Papernot}, \bibinfo{person}{Patrick
  McDaniel}, \bibinfo{person}{Somesh Jha}, \bibinfo{person}{Matt Fredrikson},
  \bibinfo{person}{Z~Berkay Celik}, {and} \bibinfo{person}{Ananthram Swami}.}
  \bibinfo{year}{2016}\natexlab{c}.
\newblock \showarticletitle{{The Limitations of Deep Learning in Adversarial
  Settings}}. In \bibinfo{booktitle}{{\em Proceedings of the 1st IEEE European
  Symposium in Security and Privacy (EuroS\&P)}}.
\newblock


\bibitem[\protect\citeauthoryear{Papernot, McDaniel, Sinha, and
  Wellman}{Papernot et~al\mbox{.}}{2016d}]%
        {papernot2016towards}
\bibfield{author}{\bibinfo{person}{Nicolas Papernot}, \bibinfo{person}{Patrick
  McDaniel}, \bibinfo{person}{Arunesh Sinha}, {and} \bibinfo{person}{Michael
  Wellman}.} \bibinfo{year}{2016}\natexlab{d}.
\newblock \showarticletitle{Towards the Science of Security and Privacy in
  Machine Learning}.
\newblock \bibinfo{journal}{{\em arXiv preprint arXiv:1611.03814\/}}
  (\bibinfo{year}{2016}).
\newblock


\bibitem[\protect\citeauthoryear{Papernot, McDaniel, Wu, Jha, and
  Swami}{Papernot et~al\mbox{.}}{2015}]%
        {papernot2016distillation}
\bibfield{author}{\bibinfo{person}{Nicolas Papernot}, \bibinfo{person}{Patrick
  McDaniel}, \bibinfo{person}{Xi Wu}, \bibinfo{person}{Somesh Jha}, {and}
  \bibinfo{person}{Ananthram Swami}.} \bibinfo{year}{2015}\natexlab{}.
\newblock \showarticletitle{Distillation as a Defense to Adversarial
  Perturbations against Deep Neural Networks}.
\newblock \bibinfo{journal}{{\em Proceedings of the 37th IEEE Symposium on
  Security and Privacy (S\&P)\/}} (\bibinfo{year}{2015}).
\newblock


\bibitem[\protect\citeauthoryear{Reuben~Feinman}{Reuben~Feinman}{2017}]%
        {2017FeinmanDASfA}
\bibfield{author}{\bibinfo{person}{Saurabh Shintre Andrew B.~Gardner
  Reuben~Feinman, Ryan R.~Curtin}.} \bibinfo{year}{2017}\natexlab{}.
\newblock \showarticletitle{Detecting Adversarial Samples from Artifacts}.
\newblock \bibinfo{journal}{{\em CoRR\/}}  \bibinfo{volume}{abs/1703.00410}
  (\bibinfo{year}{2017}).
\newblock
\showURL{%
\url{http://arxiv.org/abs/1703.00410}}


\bibitem[\protect\citeauthoryear{Rosenbaum}{Rosenbaum}{}]%
        {Rosenbaum_anexact}
\bibfield{author}{\bibinfo{person}{Paul~R. Rosenbaum}.}
\newblock \showarticletitle{An exact distribution-free test comparing two
  multivariate distributions based on adjacency}.
\newblock \bibinfo{journal}{{\em Journal of the Royal Statistical Society B\/}}
  (\bibinfo{year}{????}), \bibinfo{pages}{2005}.
\newblock


\bibitem[\protect\citeauthoryear{Shen, Tople, and Saxena}{Shen
  et~al\mbox{.}}{2016}]%
        {DBLP:conf/acsac/ShenTS16}
\bibfield{author}{\bibinfo{person}{Shiqi Shen}, \bibinfo{person}{Shruti Tople},
  {and} \bibinfo{person}{Prateek Saxena}.} \bibinfo{year}{2016}\natexlab{}.
\newblock \showarticletitle{Auror: defending against poisoning attacks in
  collaborative deep learning systems}. In \bibinfo{booktitle}{{\em Proceedings
  of the 32nd Annual Conference on Computer Security Applications, {ACSAC}
  2016, Los Angeles, CA, USA, December 5-9, 2016}}. \bibinfo{pages}{508--519}.
\newblock
\showURL{%
\url{http://dl.acm.org/citation.cfm?id=2991125}}


\bibitem[\protect\citeauthoryear{Srndic and Laskov}{Srndic and Laskov}{2014}]%
        {DBLP:conf/sp/SrndicL14}
\bibfield{author}{\bibinfo{person}{Nedim Srndic} {and} \bibinfo{person}{Pavel
  Laskov}.} \bibinfo{year}{2014}\natexlab{}.
\newblock \showarticletitle{Practical Evasion of a Learning-Based Classifier:
  {A} Case Study}. In \bibinfo{booktitle}{{\em 2014 {IEEE} Symposium on
  Security and Privacy, {SP} 2014, Berkeley, CA, USA, May 18-21, 2014}}.
  \bibinfo{pages}{197--211}.
\newblock
\showDOI{%
\url{https://doi.org/10.1109/SP.2014.20}}


\bibitem[\protect\citeauthoryear{Szegedy, Zaremba, Sutskever, Bruna, Erhan,
  Goodfellow, and Fergus}{Szegedy et~al\mbox{.}}{2014}]%
        {szegedy2013intriguing}
\bibfield{author}{\bibinfo{person}{Christian Szegedy},
  \bibinfo{person}{Wojciech Zaremba}, \bibinfo{person}{Ilya Sutskever},
  \bibinfo{person}{Joan Bruna}, \bibinfo{person}{Dumitru Erhan},
  \bibinfo{person}{Ian Goodfellow}, {and} \bibinfo{person}{Rob Fergus}.}
  \bibinfo{year}{2014}\natexlab{}.
\newblock \showarticletitle{Intriguing properties of neural networks}. In
  \bibinfo{booktitle}{{\em Proceedings of the 2014 International Conference on
  Learning Representations}}. Computational and Biological Learning Society.
\newblock


\bibitem[\protect\citeauthoryear{Sz\'{e}kely and Rizzo}{Sz\'{e}kely and
  Rizzo}{2013}]%
        {Szekely20131249}
\bibfield{author}{\bibinfo{person}{G\'{a}bor~J. Sz\'{e}kely} {and}
  \bibinfo{person}{Maria~L. Rizzo}.} \bibinfo{year}{2013}\natexlab{}.
\newblock \showarticletitle{Energy statistics: A class of statistics based on
  distances}.
\newblock \bibinfo{journal}{{\em Journal of Statistical Planning and
  Inference\/}} \bibinfo{volume}{143}, \bibinfo{number}{8}
  (\bibinfo{year}{2013}), \bibinfo{pages}{1249 -- 1272}.
\newblock
\showISSN{0378-3758}
\showDOI{%
\url{https://doi.org/10.1016/j.jspi.2013.03.018}}


\bibitem[\protect\citeauthoryear{Wang, Gao, and Qi}{Wang et~al\mbox{.}}{2016}]%
        {DBLP:journals/corr/WangGQ16}
\bibfield{author}{\bibinfo{person}{Beilun Wang}, \bibinfo{person}{Ji Gao},
  {and} \bibinfo{person}{Yanjun Qi}.} \bibinfo{year}{2016}\natexlab{}.
\newblock \showarticletitle{A Theoretical Framework for Robustness of (Deep)
  Classifiers Under Adversarial Noise}.
\newblock \bibinfo{journal}{{\em CoRR\/}}  \bibinfo{volume}{abs/1612.00334}
  (\bibinfo{year}{2016}).
\newblock
\showURL{%
\url{http://arxiv.org/abs/1612.00334}}


\bibitem[\protect\citeauthoryear{Xu, Qi, and Evans}{Xu et~al\mbox{.}}{2016}]%
        {xu2016automatically}
\bibfield{author}{\bibinfo{person}{Weilin Xu}, \bibinfo{person}{Yanjun Qi},
  {and} \bibinfo{person}{David Evans}.} \bibinfo{year}{2016}\natexlab{}.
\newblock \showarticletitle{Automatically evading classifiers}. In
  \bibinfo{booktitle}{{\em Proceedings of the 2016 Network and Distributed
  Systems Symposium}}.
\newblock


\end{thebibliography}

\end{document}